\documentclass[smallextended]{svjour3}

\usepackage[utf8]{inputenc}
\usepackage{mathpazo}
\usepackage{amsmath}
\usepackage[numbers]{natbib}
\usepackage[margin=1in]{geometry}
\usepackage{amssymb}
\usepackage{standalone}
\usepackage{graphicx}
\usepackage{epstopdf}
\usepackage[caption=false]{subfig}
\usepackage[normalem]{ulem}
\usepackage[rgb,dvipsnames]{xcolor}
\usepackage{tikz}
\usepackage{bm}
\usepackage[caption=false]{subfig}
\usepackage{booktabs}
\usepackage[math]{cellspace}
\AtBeginDocument{
    \heavyrulewidth=.08em
    \lightrulewidth=.05em
    \cmidrulewidth=.03em
    \belowrulesep=.65ex
    \belowbottomsep=0pt
    \aboverulesep=.4ex
    \abovetopsep=0pt
    \cmidrulesep=\doublerulesep
    \cmidrulekern=.5em
    \defaultaddspace=.5em
}

\definecolor{MyBlue}{rgb}  {0.1,0.1,0.9}
\definecolor{MyRed}{rgb}   {0.9,0.1,0.1}
\definecolor{MyGreen}{rgb} {0.05,0.4,0.05}
\definecolor{burntorange}{rgb}{0.8, 0.33, 0.0}
\definecolor{NeilMagenta}{rgb}{0.8, 0.1, 0.8}

\newcommand{\dt}[1]{\frac{\partial #1 }{\partial t}}

\newcommand \beq{\begin{eqnarray}}
\newcommand \eeq{\end{eqnarray}}
\newcommand \beqno{\begin{eqnarray*}}
\newcommand \eeqno{\end{eqnarray*}}
\newcommand \bit{\begin{itemize}}
\newcommand \eit{\end{itemize}}

\newcommand \nn{\nonumber} 

\begin{document}
\title{Isolating Patterns in Open Reaction-Diffusion Systems}
\titlerunning{Isolating Patterns in Open Reaction-Diffusion Systems}
\author{Andrew L. Krause \and V\'{a}clav Klika \and Philip K. Maini \and Denis Headon  \and Eamonn A. Gaffney}
\authorrunning{A. L. Krause \and V. Klika \and P. K. Maini \and D. Headon \and E. A. Gaffney}

\institute{A. L. Krause \and P. K. Maini \and E. A. Gaffney 
    \at Wolfson Centre for Mathematical Biology, Mathematical Institute, University of Oxford, Andrew Wiles Building, Radcliffe Observatory Quarter, Woodstock Road, Oxford, OX2 6GG, United Kingdom 
\and
V. Klika 
    \at Department of Mathematics, FNSPE, Czech Technical University in Prague, Trojanova 13, 120 00 Praha, Czech Republic
\and
D. Headon 
    \at The Roslin Institute and Royal (Dick) School of Veterinary Studies, University of Edinburgh, Easter Bush Campus, Midlothian EH25 9RG, United Kingdom}

\date{Received: date / Accepted: date}

\maketitle

\begin{abstract}
Realistic examples of reaction-diffusion phenomena governing spatial and spatiotemporal pattern formation are rarely isolated systems, either chemically or thermodynamically. However, even formulations of `open' reaction-diffusion systems often neglect the role of domain boundaries. Most idealizations of closed reaction-diffusion systems employ  no-flux boundary conditions, and often patterns will form up to, or along, these boundaries. Motivated by boundaries of patterning fields related to the emergence of spatial form in embryonic development, we propose a set of mixed boundary conditions for a two-species reaction-diffusion system which forms inhomogeneous solutions away from the boundary of the domain for a variety of different reaction kinetics, with a prescribed uniform state near the boundary. We show that these boundary conditions can be derived from a larger heterogeneous field, indicating that these conditions can arise naturally if cell signalling or other properties of the medium vary in space. We explain the basic mechanisms behind this pattern localization, and demonstrate that it can capture a large range of localized patterning in one, two, and three dimensions, and that this framework can be applied to systems involving more than two species. Furthermore, the boundary conditions proposed lead to more symmetrical patterns on the interior of the domain, and plausibly capture more realistic boundaries in developmental systems. Finally, we show that these isolated patterns are more robust to fluctuations in initial conditions, and that they allow intriguing possibilities of pattern selection via geometry, distinct from known selection mechanisms.
\end{abstract}
 

\maketitle


\section{Introduction}

Reaction-diffusion systems (RDS) are employed to model an increasingly wide-range of phenomena \cite{ball2001self, kuramoto2003chemical, murray2004mathematical}. Following Turing's work \cite{turing1952chemical}, an enormous literature has developed using these models to capture aspects of patterns seen in biology and chemistry \cite{gierer1972theory, murray1981pre, de1991turing, cross1993pattern, kondo2010reaction, maini2012turing}. However, as noted by Turing, these models are idealizations, and cannot account for many observations of real developmental systems \cite{maini2012turing}. We further extend present efforts to increase the realism, and hence the explanatory power, of this theory by considering the role of boundary conditions in isolating patterns away from domain boundaries. By `pattern' we mean a localized solution (a non-random spatial arrangement) of a morphogen above a threshold value, which in embryonic settings would correspond to a sufficient level to induce biologically meaningful change in cell state. Our main aim here is to demonstrate that a simple choice of boundary conditions can guarantee interior localization of patterns. By deriving these boundary conditions from a heterogeneous problem, we argue that such pattern isolation can be understood as arising from different kinds of boundaries due to differential gene expression across a field, with patterns forming only in some sub-region of the domain.

Stationary Turing patterns, as well as other complex behaviours of RDS such as oscillations and chemical chaos, are genuinely non-equilibrium thermodynamic processes. Since the early work of Nicolis and Prigogine \cite{prigogine1971biological, nicolis1977self}, substantial further work has extended this thermodynamic perspective of Turing instabilities as a non-equilibrium phenomenon associated with open systems \cite{cross1993pattern, Ross2008, falasco2018information, esposito2020open}. However, very little emphasis in these works is put on the role of boundary conditions. In the classical review of pattern selection \citep{borckmans1995turing}, for instance, the authors explicitly state that they will not be concerned with the role of boundary conditions. While many authors describe the impact of boundary conditions on such non-equilibrium phenomena \citep{cross1993pattern, murray2004mathematical}, relatively few consider more exotic conditions beyond the standard Neumann, Dirichlet, or periodic settings \citep{setayeshgar1998turing, dillon1994pattern, klika2018domain}. Physically, periodic boundary conditions can be justified for a flat approximation of a curved manifold, which can be appropriate for patterns appearing across the entire surface of an organism. However, the typical Neumann boundary conditions used to study pattern formation are very clearly a mathematical idealization of what are often open systems. Arcuri and Murray \cite{arcuri1986pattern} have suggested that such idealized boundaries are likely unrealistic, and that inhomogeneous boundary conditions can lead to more robust pattern formation (i.e.~less sensitivity on initial conditions).

Many developmental systems exhibit localized patterning within a structurally homogeneous field (i.e.~a domain without any spatial heterogeneity in diffusive fluxes or reactions due to variations in cell type or arrangement). Examples include ectodermal structures such as hair or teeth \cite{tucker2004cutting, johansson2014regionalisation}, and the role of auxin in plant root initiation \cite{duckett1994clonal, fischer2006vectorial, avitabile2018spot}. Periodic patterning of the primary hair follicles\footnote{Claxton and others have noted that in mature skin of many species of mammals, the hair follicles are present in clusters and are not evenly periodic. This arises from there being different waves of hair follicle formation, with the first of these being truly periodic and the one that relates to the breaking of symmetry \cite{claxton1964determination, nagorcka1985role, glover2017hierarchical}. The subsequent waves then are influenced by the primary hair follicles of the first wave and yield numerous follicles that cluster around the primary follicles.} in mammalian whiskers and hair more generally, for instance, suggests a regionalization of different patterning fields within which RDS may explain the emergence of patterns. However, as noted in \cite{murray2004mathematical}, Neumann boundary conditions can exhibit patterning all the way up to the boundary, often leading to partial patterns at the boundary (e.g.~half-spots in the case of spotty patterns). Of course, inhomogeneous solutions of nonlinear RDS may exhibit a variety of different behaviours, depending on the parameters and initial data of the system. In many cases (such as in the limit of large diffusivity ratios) it has been shown that spike solutions can approach, and be pinned to, boundaries, particularly points of extremal curvature in multiple spatial dimensions \cite{iron2000metastable, iron2000dynamics, kolokolnikov2004bifurcation, miyamoto2005stability, ei2013dynamics}. In other cases, initial data consisting only of internal spots can be shown to remain in a configuration of internal spots due to boundary repulsion \cite{kolokolnikov2009spot, chen2011stability}. Numerically, as we will show in Section \ref{numerics_section}, Neumann conditions will generically allow patterns to form up to the boundary for a variety of systems and parameters starting from small perturbations of a homogeneous steady state.

There have been several different approaches to designing RDS which exhibit patterns isolated away from the boundaries. A simple approach pursued by \cite{varea1997confined} was to modify the reaction kinetics at a boundary to push the system outside the Turing regime locally, so that patterning was restricted to an interior region. Similar ideas were explored with a variety of more complex spatial heterogeneities in reaction kinetics \cite{page2003pattern, page2005complex}, or diffusive fluxes \cite{benson1993diffusion}. Recently, this kind of \emph{localized} patterning has been justified in the linear regime either in the case of smoothly varying heterogeneity \cite{krause_WKB} or jumps (step functions) in the kinetics \cite{kozak2019pattern}. Similar work has been considered in some nonlinear regimes, where spike pinning, and hence pattern localization, can be obtained via heterogeneity in kinetics \cite{ward2002dynamics, avitabile2018spot}.  Another approach, pursued in the context of the positioning of bacterial protein clusters, is to understand pattern selection from a nonlinear theory, and look for nonlinearities which exhibit the desired isolated patterns \cite{murray2017self, subramanian2020pattern}. 

More generally, several studies have investigated inhomogeneous or mixed boundary conditions. Dirichlet boundary conditions can have a variety of effects on patterns, including modifying parameters for which patterns are observed \cite{maini1997boundary}. Robin conditions in pattern-forming RDS have recently been shown to impact the number of interior spots \cite{tzou2011stationary}, and inhomogeneous fluxes of inhibitor can lead to movement of interior pulse solutions in 1-D towards the source of the flux (in addition to changing amplitudes and spacing) \cite{tzou2018stability}. Closer to what we will propose here, \cite{dillon1994pattern} investigated a variety of different boundary conditions for each species in a two-species RDS, demonstrating that different pattern selection mechanisms could be influenced by the choice of these boundary conditions.

While we are primarily interested in understanding isolated patterns, our work also falls into the larger context of trying to connect Turing's simple patterning mechanism with the complex reality of biological development. As Turing himself said, pattern formation occurs in stages through subsequent patterning \cite{turing1952chemical}. There are difficult questions of how to idealize such situations to elucidate the fundamental mechanisms at play in any particular stage. One such question is the robustness of Turing patterns, both in terms of the dependence of steady patterned states on initial data, and the strict requirements on parameters needed for RDS to admit Turing-type patterns \cite{maini2012turing, woolley2017turing, scholes2019comprehensive}. Domain growth has been suggested as one way to help improve robustness \cite{crampin2002pattern, krause2019influence, van_gorder_growth_2019}, as has stochasticity \cite{woolley2011stochastic}. Here, we will suggest that RDS exhibiting isolated patterning will also be more robust in both enlarging the parameter space within which we see patterns, as discussed in \cite{maini1997boundary}, and in admitting fewer possible steady state solutions.


In developmental settings, the source of boundaries between regions which exhibit periodic patterning, and those which do not, can vary between tissue type and the specific morphogen signalling dynamics. In particular, boundaries can arise due to either explicit heterogeneity in the tissue (where cells of different types explicitly demarcate patterning regions, due to previous fate determination) or due to more complex and diffuse mechanisms involving cell state, which may be refined into sharp boundaries via, e.g., bistability. Such boundaries can be thought of as an example of positional information \cite{meinhardt1983boundary, green2015positional}. We briefly discuss the biology of such boundary formation before presenting our modelling framework. 

\subsection{Biological Interlude: Boundaries in Developmental Patterning}\label{DevSubsect}
Activator and inhibitor species are taken to represent specific molecules (or more accurately, signalling pathways) in biological systems. These undergo processes of synthesis, decay, and diffusion, depending on the type of molecule employed as a signal. Most intercellular communication is achieved through gene-encoded polypeptides (proteins), with some smaller molecules also playing roles as diffusible signals. The synthesis of these signals is largely controlled by rates of gene expression (transcription and translation) within cells, directly in the case of proteins, and for small molecules indirectly through production of enzymes that catalyse their formation. These molecules are secreted from cells, then diffuse, either alone or interacting with other proteins, through subcellular structures, or the meshwork of extracellular matrix. Ultimately, these molecules are capable of attaching to receptors on the surface of, or sometimes within, cells in the vicinity. The effect of receptor binding is to trigger a typically multi-step process of signal transduction, ultimately altering rates of gene expression (molecule synthesis), or cell behaviour (such as cell shape or movement) \cite{bradshaw2009handbook}.

Many extracellular proteins that act as signals are themselves bound in the extracellular space by inhibitory or transport proteins. Thus inactive complexes can be formed, capable of diffusing but unable to bind receptors and elicit a response. An example is the Bone Morphogenetic Protein (BMP) family, several members of which have been implicated in reaction-diffusion periodic patterning systems of skin \cite{ho2019feather, glover2017hierarchical}, limb skeleton \cite{raspopovic2014digit}, and gut \cite{walton2016villification}, and for which a range of distinct extracellular inhibitor proteins have been characterised \cite{walsh2010extracellular}. 

Two points relevant to this work that arise from these general features of cell-cell communication are i) that the specificity of molecular interactions is very high, and ii) that the candidate activator and inhibitor molecules in many systems do not interact directly, but rather through chains of intermediates in signalling pathways. The specificity of molecular interactions and indirect nature of activator-inhibitor interaction permit selective spatial variation in the properties of the activator and inhibitor that we will explore in this paper. Boundaries relevant to reaction-diffusion patterning will be defined by the geometry of the embryo or organ, and by spatial variation in the expression of genes that encode or modulate activator and inhibitor synthesis, diffusion, biological activity, and decay. Geometric boundaries are unavoidable and often prominent in experimental systems at the cut edges of the tissue, as they have a strong influence on pattern formation \cite{glover2017hierarchical}. 

Variation in gene expression can come about due to differences in tissue structure that cause entirely distinct cell types to be directly apposed. This is observed in composite organs such as skin and gut, in which distinct and non-intermingling cell types with deeply divergent developmental origins are organised into a tightly-packed epithelial sheet and a looser mesenchyme rich in extracellular matrix. In each of these tissues the cells express a distinct subset of the genome, have different shapes and mobility, and the extracellular space between the cells has very different properties. These two tissue types are separated by a specialised planar matrix called the basement membrane. The locations of boundaries between these components are marked by abrupt changes in tissue structure that are readily visible at the anatomical or histological levels.

\begin{figure}
    \centering
    \subfloat[]{\includegraphics[width=0.6\textwidth]{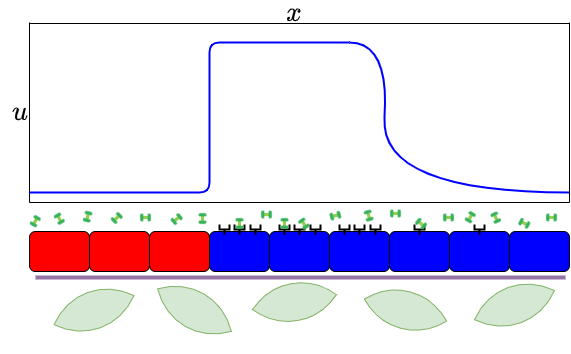}}
    \subfloat[]{\includegraphics[width=0.3\textwidth]{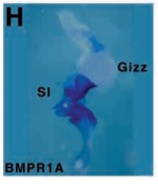}}
    \caption{(a) A graphical representation of three kinds of boundaries, leading to the concentration profile of the intracellular protein $u$ shown at the top. The production of this protein is stimulated as a result of the activation of black receptors after binding to the dumbbell-shaped extracellular signalling molecules. The red and blue rectangular cells represent two distinct cell states, with the blue capable of expressing the receptor. The green oblate cells instead represent a different cell type, separated by a basement membrane (which, in our depiction, is assumed to not influence the intracellular concentration of $u$). Finally, the blue cells also have a gradient of receptor expression towards the right boundary (which itself may be a hard no-flux boundary), leading to a more diffuse intracellular concentration of the protein $u$. (b) In situ hybridization of a chick gut (developmental stage E4.5) with a riboprobe to BMPR1A from Fig.~1(H) of \cite{smith2000roles}. The purple/blue stain shows the ability of the tissue to receive BMP signals, whereas the white/translucent areas lack this receptor and do not have the same response to BMP (though BMP may still diffuse throughout these regions). }
    \label{DevFig}
\end{figure}

On a finer scale, differences in gene expression within a particular tissue (that is, composed primarily of cells of the same type), can arise from previous patterning events, such as an external organising region emitting a graded signal that influences part of the patterning field. Graded boundaries may persist in that form, or can be refined into sharp boundaries by, for example, the action of mutually antagonistic factors that create bistable systems, notable in the formation of discrete segments \cite{briscoe2015morphogen}. If the output of such systems influences the production of a receptor or an extracellular inhibitor for a component of the reaction-diffusion system, then across the boundary this factor would be predicted to diffuse, but to lack any biological activity. Boundaries characterised by differences in gene expression may be present in structurally homogeneous tissues and require application of specialised molecular methods to reveal them. For example, in vertebrate embryonic gut regionalization, initially diffuse boundaries become sharper \cite{li2009dynamic}, and BMP receptor expression becomes distinct in different regions or segments along the tract \cite{smith2000roles}. In Figure \ref{DevFig}(a) we depict both kinds of boundaries, and in (b) we show an example of a boundary observed in the ability of a tissue to respond to BMP within the otherwise-homogeneous gut tract described above.

Thus, the spatially varying presence or absence of a receptor, a component of its signal transduction pathway, or an extracellular inhibitor, can itself constitute a boundary relevant to the behaviour of a reaction-diffusion patterning system, even in a structurally homogeneous field of cells. Such a boundary can have a selective effect on a single species (activator or inhibitor) in the reaction-diffusion system, which will be free to diffuse across the boundary and persist physically, but without the ability to influence reaction kinetics in this region of the field.

\subsection{Mixed Boundary Conditions}\label{intro_model}

We now consider modelling such boundaries in a reaction-diffusion system. Throughout the paper we will consider a generic two-species RDS in the domain $x \in \Omega \subset \mathbb{R}^n$:
\begin{equation}\label{ueq}
    \dt{u} = \nabla^2 u + f(u,v),
\end{equation}
\begin{equation}\label{veq}
    \dt{v} = D \nabla^2 v + g(u,v),
\end{equation}
where $D>1$ is a ratio of diffusion coefficients. In the classical case of activator-inhibitor kinetics, where $v$ is a self-inhibitor, $D>1$ is a necessary condition for inhomogeneous steady solutions (at least in convex domains \citep{kishimoto1985spatial}). Hereafter we restrict attention to this case in our analysis, and consider $u$ and $v$ as presumptive activator and inhibitor, respectively.

We will show that inhomogeneous steady solutions (Turing-type patterns and otherwise) can be isolated away from the domain boundary. This can be achieved by using the boundary conditions
\begin{equation}\label{mixed_BCs}
    u=R, \quad \bm{n}\cdot \nabla v=0, \quad \textrm{for all }\bm{x} \in \partial \Omega,
\end{equation}
where $R$ should be approximately a minimum of $u$ in a patterned state (e.g.~with Neumann boundary conditions). Specifically, if we compute a stable steady state solution given Neumann boundary conditions and initial data perturbed from the homogeneous equilibrium, we will take $R=\min_\Omega(u)$. Below we will show that the precise value of $R$ seems to play only a small role in influencing the pattern, at least for some reaction kinetics. While it can have quantitative effects on the size and spacing of localized solutions, our numerical results suggest that taking $R$ less than the homogeneous steady state suffices to lead to interior patterning. In some sense this is obvious, as we are fixing the value of $u$ at the boundary. We will focus on the case where the activator is fixed at the boundary and the inhibitor is not allowed to diffuse (i.e.~satisfies zero flux conditions). In principle our results on interior localization will hold if we swap the boundary conditions between activator and inhibitor, but the lengthscale of the boundary influence will be larger in the case of the inhibitor being fixed (as $D>1$, and often $D\gg 1$, for a pattern-forming RDS). Finally we note that other values of $R$ may be suitable for the development of interior localization, particularly in the presence of multistability or multiple species interactions; see Section \ref{many_species} for examples.

In developmental settings, the discussion in the preceding subsection allows us to consider $u$ and $v$ as behaving differently outside the domain $\Omega$ due to boundaries in receptor distribution or gene expression for example; in turn this can lead to different boundary conditions for the activator and inhibitor. We will justify the boundary conditions given in \eqref{mixed_BCs} asymptotically in Section \ref{asymptotic_sec} by explicitly considering a spatially heterogeneous extension of the system defined on a larger domain. We also remark that these conditions arise naturally in the case that $u$ represents temperature, and $v$ a chemical which undergoes exothermic or endothermic reactions \cite{serna2017thermodynamic, van_gorderthermal2020}. Throughout we will refer to \eqref{mixed_BCs} as mixed boundary conditions, though other authors have used the term nonscalar boundary conditions \cite{dillon1994pattern}. We will contrast these boundary conditions with homogeneous Neumann conditions on both species, and refer to the latter simply as Neumann conditions for brevity throughout the manuscript.

The rest of the paper is organized as follows. In Section \ref{explanation_section} we analytically show how the mixed boundary conditions \eqref{mixed_BCs} can be obtained from a heterogeneous problem on an enlarged domain. We also describe how these boundary conditions force pattern isolation away from the boundary in terms of a local pattern-selection mechanism. In Section \ref{numerics_section}, we demonstrate a broad numerical exploration of this interior localization across a variety of two-species reaction kinetics with qualitatively distinct kinds of patterns. We also show that the same mechanism can operate in three spatial dimensions, as well as in systems with more than two species. Finally, in Section \ref{discussion_section} we conclude by discussing both phenomenological and mechanistic applications of these boundary conditions, as well as a number of further research directions.

\section{Heterogeneity and Isolated Patterning}\label{explanation_section}

We now explore two aspects of the boundary conditions \eqref{mixed_BCs}. First we explain how they arise from an asymptotic analysis of a heterogeneous problem, motivated by regions of different gene expression as discussed in Section \ref{DevSubsect}. Viewing these conditions as arising from spatial heterogeneity is consistent both with the biological observations discussed there, as well as with the theoretical literature on localization of patterns in heterogeneous media. We then discuss steady state selection mechanisms observed from different boundary conditions, explaining some aspects of the isolated patterning we expect to achieve with these boundary conditions.

\subsection{Derivation of Mixed Boundary Conditions from a Heterogeneous Problem}\label{asymptotic_sec}

{Working   non-dimensionally throughout, we} consider a larger domain $\tilde{\Omega} \subset \mathbb{R}^n$, and a strict subset of it ${\Omega}\subset \tilde{\Omega}$ such that $\tilde{\Omega} \setminus {\Omega}$ is connected, with all domain boundaries sufficiently smooth. We now consider the non-dimensional heterogeneous reaction-diffusion system,
\begin{equation}\label{ueq_het}
    \dt{u} = \nabla^2 u + \begin{cases}
    f(u,v), \quad &\textrm{for } \bm{x} \in {\Omega},\\
    \rho(R- u), \quad &\textrm{for } \bm{x} \in \tilde{\Omega} \setminus {\Omega},
    \end{cases}
\end{equation}
\begin{equation}\label{veq_het}
    \dt{v} = D \nabla^2 v + \begin{cases}
    g(u,v), \quad &\textrm{for } \bm{x} \in {\Omega},\\
    g_2(u,v), \quad &\textrm{for } \bm{x} \in \tilde{\Omega} \setminus {\Omega}.

    \end{cases}
\end{equation}
For concreteness we consider Neumann conditions for both species on $\partial \tilde{\Omega}$.

This heterogeneous system can be understood as two regions of different activity (e.g.~signal transduction and gene expression) in the medium. Within the interior region, ${\Omega}$, the normal activator-inhibitor dynamics are present, whereas in the exterior region, $\tilde{\Omega} \setminus {\Omega}$, 
the activator relaxes to an equilibrium concentration on a { non-dimensional} timescale of $1/\rho$,  while the inhibitor may interact with the activator in a different manner $g_2$ with a timescale $T_{g_2}$. Such dynamics are easily obtained in examples where gene expression is spatially modulated due to previous symmetry breaking and fate specification events, leading to heterogeneity in local signalling dynamics or in tissue/cell type as discussed in Section \ref{DevSubsect}. 

One can show that for sufficiently large $\rho$, the one-dimensional steady state version of Equations \eqref{ueq_het}-\eqref{veq_het} can, to leading asymptotic order, be reduced to considering the steady state problem \eqref{ueq}-\eqref{veq} on the smaller domain $\Omega$ with the boundary conditions \eqref{mixed_BCs}. In higher dimensions, however, one must impose additional geometric constraints. For concreteness, to capture the general case we consider the problem in two spatial dimensions, remarking that extensions to higher dimensions follow the same reasoning. We let $\Gamma = \partial \Omega$ be the boundary between the internal and external regions of $\tilde{\Omega}$ (equivalently, the internal boundary of $\tilde{\Omega}\setminus \Omega$), and define $\eta$ as the (non-dimensional) ratio of the maximum thickness (normal to $\Gamma$) of $\tilde{\Omega}\setminus\Omega$ to the arclength of $\Gamma$. The one-dimensional case follows exactly the same reasoning as given below for two dimensions, without the technicalities related to the geometry.

\begin{figure}
    \centering
    \includegraphics[width=0.7\textwidth]{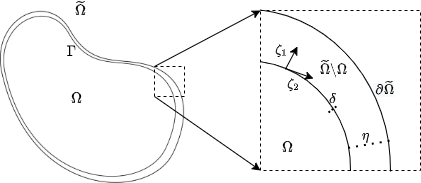}
    \caption{The geometry of the full domain $\tilde{\Omega}$, with a region in $\tilde{\Omega}\setminus \Omega$ enlarged to show the various asymptotic parameters and coordinates along the boundary discussed in the text. Note that $\eta$ is, in general, a ratio of the maximum thickness of $\tilde{\Omega}\setminus\Omega$ to arclength of $\Gamma$, but for illustrative purposes we assume the arclength is order unity. Similarly we assume sufficient smoothness of both internal and exterior boundaries so that the local mean curvature is always order unity or smaller. 
    }
    \label{BoundaryFig}
\end{figure}

We proceed to  show that for sufficiently large $\rho $ and for $\tilde{\Omega} \setminus {\Omega}$ sufficiently slender and regular, steady states of this system asymptotically reduce to those of the reaction-diffusion system \eqref{ueq}-\eqref{mixed_BCs} defined on the interior domain $\Omega$. Additionally, the dynamics of $u$ and $v$ outside the interior domain are simple and (asymptotically) determined analytically.  In particular, with $\eta \ll 1$ denoting the assumption of slenderness for $\tilde{\Omega} \setminus {\Omega}$, we consider the asymptotic regime $\rho \gg 1$ and $\eta \ll 1$. We focus on steady states, as transient solutions to the heterogeneous problem \eqref{ueq_het}-\eqref{veq_het} need not satisfy the no-flux boundary condition for $v$ on $\partial \Omega$ in this asymptotic scaling without further restrictions on kinetic timescales. Our results will then show that the steady states of \eqref{ueq}-\eqref{mixed_BCs} will be precisely the same as those of \eqref{ueq_het}-\eqref{veq_het}. Similarly we will need to exploit the geometry of a slender domain to prevent problems with transverse gradients in $v$; in one spatial dimension the restriction of $\eta \ll 1$ can be relaxed completely. See Figure \ref{BoundaryFig} for a geometric depiction.

Our first objective is to determine how the solution within the interior of the boundary of $\Gamma$  is related to the solution   in the exterior of $\Gamma$ on approaching the boundary $\Gamma$. 
To proceed, we first mollify the transition between the two regions with a function $H_\delta$ which is centred on this boundary and transitions from zero to unity in the normal direction of the inner boundary of $\tilde{\Omega} \setminus {\Omega}$. We also assume $\Gamma$ has a well-defined and bounded curvature. Furthermore the transition across this boundary is on a (dimensionless) lengthscale of  $\delta \ll  \eta $,  and since the mollification scale $\delta$ is for mathematical convenience it is taken  to be smaller than any physical scale below.

The mollified steady state equations
  on the extended domain are 
\begin{equation}\label{u1}
    0 = \nabla^2 u +  H_\delta 
    f(u,v) +(1-H_\delta) \rho(R- u) , ~~~~~~~~
    0 = D \nabla^2 v +   H_\delta 
    g(u,v) . 
\end{equation}
We introduce an orthonormal curvilinear
 coordinate system given by $  (   \zeta _{1  },  \zeta _{2  }) $ along the curve $\Gamma$, 
where the level set $  \zeta_{1  }  =0 $ corresponds to $\Gamma$ with  the $  \zeta_{1  }   $ axis
 perpendicular to $\Gamma$  and pointing into the interior of  $\tilde{\Omega} \setminus {\Omega}$. We now follow the derivation of an Eikonal equation along the boundary (see, e.g., \cite{keener1998mathematical}). Using the Einstein summation convention, the chain rule gives 
\beq  \nn 
\frac{\partial    }{ \partial x_i  } & = &\frac{\partial  \zeta _{j  }     }{\partial x_{i  }    }
\frac{\partial    }{  \partial  \zeta _{j  }  }   =: \alpha_{ij} 
\frac{\partial    }{  \partial  \zeta _{j  }  } 
\eeq 
and hence, for example, 
\beq \nn 
 \nabla^2 v & = & 
  \alpha_{ip  } \alpha_{iq  } \frac{\partial^{2  } v      }{\partial  \zeta _p \partial  \zeta _{q  }  }  
+ 
\frac{\partial \alpha_{ip  }     }{ \partial x_{i  }   } \frac{\partial v    }{ \partial  \zeta _{p  }   } .
\eeq
We have that the vector 
\beq   \nonumber \nabla_{x} \zeta_1=
\left(\frac{ \partial \zeta_1    }{  \partial x_{1  }} ,  \frac{\partial \zeta_1    }{  \partial x_{2  }  } \right) =  
 (\alpha_{11}, \alpha_{21} ) =  
\mbox{\boldmath{$\alpha$}},
\eeq
is normal to $\Gamma$ and, by the need for orthonormality,  we fix the scale of $\zeta_1$    so that $\mbox{\boldmath{$\alpha$}}$ is a unit normal and then   $\kappa := \nabla\cdot \mbox{\boldmath{$\alpha$}}$ is the mean curvature of $\Gamma$. In particular assuming the curvature is of order unity relative to the small parameter $\delta$ entails the scale of $\zeta_1$ is order unity. Further noting 
$$ \alpha_{i1}\alpha_{i2} = \nabla_x\zeta_1 \cdot \nabla_x\zeta_2=0,$$ by orthogonality of the $(\zeta_1,\zeta_2)$ coordinate system, we have 
\beq \nn 
 \nabla^2 v & = & 
 \frac{\partial^{2  }   v  }{\partial^2  \zeta _1   } 
+  \kappa  
\frac{\partial v    }{\partial  \zeta _1   } + c_1\frac{\partial v    }{\partial \zeta _2}+c_2\frac{\partial^2 v    }{\partial \zeta _2^2},
\eeq
where the coefficients $c_1,c_2$ come from the Jacobian of the transformation and are of the same order as $\kappa$ or smaller. Hence, these terms will be asymptotically small once we rescale $\zeta_1$ below. Derivatives of $u$ will have an identical expansion.

Given the scale of the transition of $H_\delta$, we now consider a prospective boundary layer of width of the scale $\delta$ at the boundary $\Gamma$ and thus we consider $\zeta'_1 =\zeta_1/\delta, ~\zeta'_2 =\zeta_2.  $  
Hence at leading order we have $u_{\zeta'_1\zeta'_1}=v_{\zeta'_1\zeta'_1}=0$ in the transition region and so $v$ is linear in $\zeta'_1$, i.e. { $$v=P(\zeta'_2)\zeta'_1+Q(\zeta'_2),$$}  and similarly for $u$ in this region. However boundedness of this inner solution when leaving the transition region (i.e.~as $\zeta'_1\rightarrow\infty$) forces { $P(\zeta'_2)$} to be zero so that matching to an outer solution is possible, and thus 
$u=u(\zeta'_2)=u(\zeta_2), ~v=v(\zeta'_2)=v(\zeta_2)$. The independence of $u$ and $v$ from the normal coordinate, $\zeta'_1$, in the prospective boundary layer, means that on matching into this layer on either side of $\Gamma$, we attain continuity of $u$ and $v$.  
Within the $\delta$-boundary layer we have $\partial v/\partial \zeta'_1\sim 0$ at leading order; however,  this does not mean physical fluxes are essentially zero, since the mollification lengthscale $\delta$  is taken to be smaller than any physical lengthscale and $v$ may vary significantly over the latter, with analogous remarks for $u$. Nonetheless,   the collapse  of the prospective $\delta$-boundary layer strongly suggests  the continuity of $u,v$ across $\Gamma$ should be supplemented by continuity of flux.

To explicitly deduce  this, let $z_2$ such that $|\zeta_2-z_2|\leq \delta^{1/2}$ be fixed and consider the integral form of the steady state conservation relation (given by \eqref{u1}) on the pillbox region 
$${\cal P}=\{(\zeta_1,\zeta_2)| \zeta_1 \in [-\delta,\delta],~~\zeta_2-z_2 \in [-\delta^{1/2},\delta^{1/2}\}]. $$
Integrating the conservation equation \eqref{u1} for $v$ over the pillbox and applying Green's Theorem gives 
\beq 
0=D\int_{z_2-\delta^{1/2}}^{z_2+\delta^{1/2}} -\frac{\partial v }{\partial \zeta_1}( -\delta,\bar{\zeta}_2) \mathrm{d}\bar{\zeta}_2 +D\int_{z_2-\delta^{1/2}}^{z_2+\delta^{1/2}}  \frac{\partial v }{\partial \zeta_1}( \delta,\bar{\zeta}_2) \mathrm{d}\bar{\zeta}_2 +O(\delta)+O(\rho \delta^{3/2}),
\eeq 
and similarly for $u$. In particular, the absence of a boundary layer at $\Gamma$ entails that $u,v$ are bounded, with  bounded normal derivatives with bounds independent of $\delta$. Thus    the $O(\delta)$ terms emerge from the other flux boundary integrals and the $O(\rho \delta^{3/2})$  term  emerges from the integrals of sinks and sources over the pillbox, noting that the area of the pillbox ${\cal P}$ is  $4\delta^{3/2}$ and the largest source term scales with $\rho \gg 1$. However, the asymptotic parameter $\delta$ is from the mollification of a Heaviside function and thus we can take $\delta^{3/2}\rho \sim O(\delta)$, since $\delta$ has been introduced for mathematical convenience and can be taken as small as required, whereas $\rho$ is constrained by the underlying biophysical limits of source production in the interpretation of the model. Hence, using the integral mean value theorem on the above integrals, dividing by $\delta^{1/2}$ and taking the limit $\delta\rightarrow 0$, we have continuity of flux, and thus normal derivative, across $\Gamma$ for $v$, and analogously for $u$, as expected.

Knowing how to match across $\Gamma$ we now need to consider  our next objective, which  is to determine   the behaviour of the solutions within the region $\tilde{\Omega} \setminus {\Omega}$ to assign boundary conditions for a reduced model on the boundary $\Gamma$. We take advantage of the  slenderness of $\tilde{\Omega} \setminus {\Omega}$, characterised by an aspect ratio of its normal extent to the arclength of $\Gamma$ via $\eta \ll 1 $. 
It is most useful to see if non-trivial behaviour may occur in the normal direction  as, if present,  such behaviour  would preclude a homogeneous Neumann condition for $v$ in the reduced system.  We thus rescale  $\zeta^*_1 =\zeta_1/\eta, ~\zeta^*_2 =\zeta_2,  $ and at leading order in $\eta$ we have
$$D\left(\eta^{-2} v_{\zeta^*_1\zeta^*_1}+O\left (\eta^{-1}\right )\right)=g_2(u,v).$$
If the characteristic time of the inhibitor kinetics in the exterior, $T_{g_2}$, is large when compared to that of diffusion, $\eta^2 D^{-1}\ll T_{g_2}$, we finally have, from \eqref{u1},
$$ v_{\zeta^*_1\zeta^*_1}  =0,$$
at least once away from $\Gamma$ by a scale of more than $\delta\ll \eta$. We have also assumed that the domain of  $\tilde{\Omega} \setminus {\Omega}$ is sufficiently regular to ensure   $\kappa \sim o(1/\eta)$, which will occur if the radius of curvature is on the lengthscale of the perimeter of $\tilde{\Omega} \setminus {\Omega}$, rather than the smaller lengthscale of its thickness. 

We now need to make a further regularity assumption about $\tilde{\Omega} \setminus {\Omega}$, namely  that its external boundary  can be specified by $\zeta^*_1=F(\zeta^*_2)$ where $F\sim O(1)$   by construction, and is also 
a single-valued function that satisfies $\eta |F'(\zeta^*_2)| \ll 1, $ where prime denotes derivative. Hence we do not consider an external boundary consisting of a curve that,  for example, has a cusp or varies too  rapidly. From this assumption, we have that the normal derivative operator on the exterior boundary of $\tilde{\Omega} \setminus {\Omega}$ is proportional to 
$$\frac{\partial}{\partial \zeta^*_1}-\eta F'(\zeta^*_2)\frac{\partial}{\partial \zeta^*_2}  \approx \frac{\partial}{\partial \zeta^*_1}.$$
Thus, working at leading order in $\eta |F'(\zeta_2)| \ll 1,  $
  homogeneous Neumann conditions on the exterior boundary of $\tilde{\Omega} \setminus {\Omega}$ give $ v_{\zeta^*_1 }=0$ at the external boundary and hence we have 
$v=v(\zeta^*_2)=v(\zeta_2) $ throughout $\tilde{\Omega} \setminus {\Omega}$, except possibly $\delta$-close to $\Gamma$. 

As will emerge below, we ultimately consider the leading order behaviour when the asymptotic parameters satisfy $1/\delta \gg \rho^{1/2}\gg 1/\eta \gg 1$. The restriction $1/\delta \gg \rho^{1/2}$ entails the scale of the mollification region, $\delta$, is the smallest scale present as required since the mollification is introduced  for analytical expedience, rather than as a fundamental feature of the biophysics. The constraint $\rho^{1/2}\gg1/\eta$
ensures the source strength is sufficiently large to enforce approximately homogeneous solutions in   the region $\tilde{\Omega} \setminus {\Omega}$, as observed below. Finally $ \eta\ll 1 $ is required to render the problem to be effectively one-dimensional at leading order and enforces $\tilde{\Omega} \setminus {\Omega}$ to be slender. To proceed, we 
consider $u$ in the region $\tilde{\Omega} \setminus {\Omega}$, whereby at leading order
\beq \label{eq123} 0=u_{\zeta^*_1\zeta^*_1} + \eta^2\rho(R-u),
\eeq 
once away from $\Gamma$ by a scale of more than $\delta$.
Noting the Neumann boundary condition on the exterior boundary, Equation \eqref{eq123} gives 
\beq \label{eq1234}  u(\zeta^*_1,~\zeta^*_2) = R + A(\zeta^*_2)\cosh\left( \rho^{1/2}\eta (F(\zeta^*_2)-  \zeta^*_1)\right).
\eeq 
For $\delta$ sufficiently small,  the above solution  will only vary by an asymptotically small amount across the   $\delta$-scale mollification region, as there is  no boundary layer. Hence, as $\zeta^*_1\rightarrow 0$, this solution  is continuous with the solution on $\Omega$ as the latter approaches $\Gamma$ for the same value of $\zeta^*_2,$ and analogously for the flux. No useful information is gained by continuity of concentration. However, continuity of flux entails $\partial u /\partial \zeta_1 \sim O(1)$ as $\rho$ increases on approaching $\Gamma$, assuming $\partial u /\partial \zeta_1 $ in $\Omega$ does not blow up with increasing  $\rho$, which is consistent with numerical simulations (below) and parabolic regularity. We then have that Equation \eqref{eq1234} yields  
\beq\label{eq12345}
\left. \frac{\partial u}{\partial \zeta_1}\right|_{\zeta _1=0^+}=
\frac{\partial \zeta^*_1}{\partial \zeta_1}
\left.\frac{\partial u}{\partial \zeta^*_1}\right|_{\zeta ^*_1=0^+}= - {\rho^{1/2}} A(\zeta^*_2)\sinh(\rho^{1/2}\eta  F(\zeta^*_2)) \sim O(1), 
\eeq 
and hence 
\beq
 | A(\zeta^*_2)| \sim O\left( \frac {1 } {\rho^{1/2}}\frac 1 {\sinh(\rho^{1/2}\eta  F(\zeta^*_2)) }\right)  . 
\eeq  

We further have that  $\rho$ is sufficiently large to ensure $\rho^{1/2}  \gg 1/\eta $, and we have already taken $\eta\ll 1$. Noting that $\cosh(p)$ is monotonically increasing for $p>0$, and 
   $ F(\zeta^*_2) \geq \zeta^*_1\geq 0, ~ F(\zeta^*_2) >0$ in  $\tilde{\Omega} \setminus {\Omega}$,
we then have for solution \eqref{eq1234} that 
\beq \nonumber 
| A(\zeta^*_2)   \cosh\left( \rho^{1/2}\eta (F(\zeta^*_2)-  \zeta^*_1)\right) | \leq |A(\zeta^*_2)|\cosh\left( \rho^{1/2}\eta F(\zeta^*_2) \right) &\sim&  O\left( \frac {1 } {\rho^{1/2}} \coth(\rho^{1/2}\eta  F(\zeta^*_2)) \right) \\ &  \sim &  O\left(  \frac {1 } {\rho^{1/2}}   \right).     
\eeq
  Thus, providing $R \gg 1/ \rho^{1/2}$, we have $u=R$ is an    estimate of the solution throughout  $\tilde{\Omega} \setminus {\Omega}$  with asymptotically small relative error. We typically take $R\sim O(1)$ so that small relative error in the approximation of $u=R$ in the region $\tilde{\Omega} \setminus {\Omega}$ is assured   by our assumption of $\rho^{1/2} \gg 1/\eta \gg 1$.  In addition, if  $R\sim O(1/\rho^{1/2})$
we have $u\sim O(1/\rho^{1/2})$, so  that the estimate $u=R $ presents with asymptotically small absolute   error,   even  though the relative error is of order unity or possibly larger.

Finally, we consider the effective boundary conditions for the steady states of the reduced system \eqref{ueq}-\eqref{veq}, \begin{equation}\label{u2}
    0 = \nabla^2 u +   
    f(u,v)   , ~~~~~~~~
    0 = D \nabla^2 v +  
    g(u,v) ,
\end{equation}
on the domain $\Omega$, which has external boundary $\Gamma$  with    $\rho^{1/2} \gg 1/\eta \gg 1$  and sufficiently regular $\tilde{\Omega} \setminus {\Omega}$.  Continuity of ﬂux immediately gives homogeneous Neumann conditions for $v$ on $\Gamma$ for the reduced system, while continuity
is  inappropriate  for  $v$,  since  $v$  is  not  determined  in $\tilde{\Omega}\setminus\Omega$,  except  ultimately  via  its  coupling  to  the solution on the interior of $\Gamma$.
In contrast, we have $u=R$ to leading order in  $\tilde{\Omega}\setminus\Omega$ and thus continuity gives the Dirichlet condition $u=R$ as a boundary condition on $\Gamma$ for the reduced system. 
Note the continuity of flux across $\Gamma$ is still required if considering the full system on  $\tilde{\Omega}$, and this would be enforced by the appropriate choice of the integration degree of freedom $A(\zeta_2^*)$ in Equation \eqref{eq1234} via  use of the expression for $\partial u/\partial\zeta_1$ at $\zeta_1=0^+$ in Equation \eqref{eq12345}. In particular using 
a Neumann boundary condition for $u$ with the reduced system, 
Equation \eqref{u2} would admit solutions that do not match continuity of $u$ across the boundary. We will explore this idea further in the next subsection, referring specifically to Figure \ref{pattern_selection2} as a demonstration of this solution selection mechanism under different boundary conditions.


In summary, we have that, providing $\tilde{\Omega} \setminus {\Omega}$ is sufficiently slender and regular, and with $\rho$ sufficiently large, the behaviour of steady solutions of  Equations \eqref{ueq_het}-\eqref{veq_het} on $\tilde{\Omega}$ is given, to asymptotic accuracy, by solutions  of Equations \eqref{u2} within the domain $\Omega$ with the boundary conditions on $\Gamma$ that $u=R$ and that  the normal derivative of $v$, that is $\partial v/\partial n$, is zero.  We note that the set of steady state solutions will match between the heterogeneous and reduced problems, but solutions to the time dependent problems could allow for solution selection based on transients, leading to a disagreement between the two models. Nevertheless, we expect this to be rare or only have small effects in typical cases, which we demonstrate numerically in Section \ref{numerics_section}. Finally we also remark that the asymptotic restrictions, particularly regarding the slender geometry, can also be observed numerically to have typically little effect on stationary solution behaviours when comparing these two models, which we demonstrate from full numerical simulations in Section \ref{numerics_section} for specific nonlinear kinetics. The driving force in deriving the conditions \eqref{mixed_BCs} is the heterogeneous switching between interacting and weakly or non-interacting species, which can be understood in the context of the biology described in Section \ref{DevSubsect}.

\subsection{Interior Patterning as Steady State Selection}\label{pattern_selection_sec}

Here we further elucidate why these boundary conditions will lead to confinement of patterned states on the interior of the domain, using a particular example of reaction kinetics in a one-dimensional domain. We consider patterned steady states as subsets of all admissible ones in the periodic case, and show how specific subsets of this set are selected by different boundary conditions. The motivation here is the discussion of equivariant bifurcation theory and the impact of symmetry on solution branches in \cite{dillon1994pattern}. Specifically, we numerically show that there are typically many solutions that satisfy Neumann or periodic conditions for $u$ and $v$, but only a subset of these satisfy the mixed conditions \eqref{mixed_BCs} when $R$ is chosen to coincide with a specific minimum value of $u$ from the Neumann solutions. In particular, solutions with these mixed boundary conditions are forced to pattern away from the boundaries as they are exactly chosen to match a Neumann solution with patterning only on the interior. We numerically observe that small changes in $R$ from this value lead to small local changes at the boundary, but broadly similar solution structures overall.

\begin{figure}
    \centering
    \subfloat[Periodic]{\includegraphics[width=0.4\textwidth]{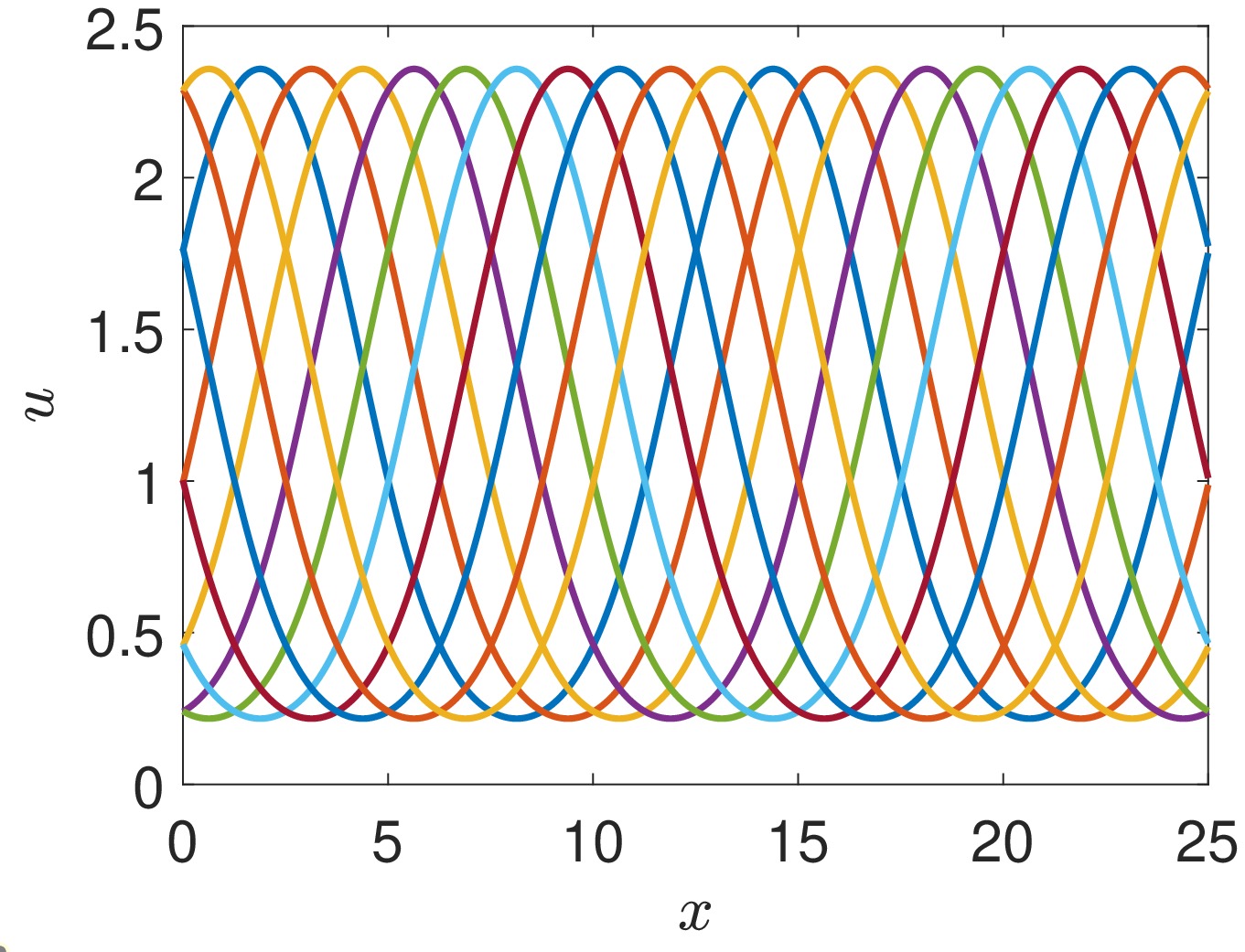}}
    \subfloat[Neumann]{\includegraphics[width=0.4\textwidth]{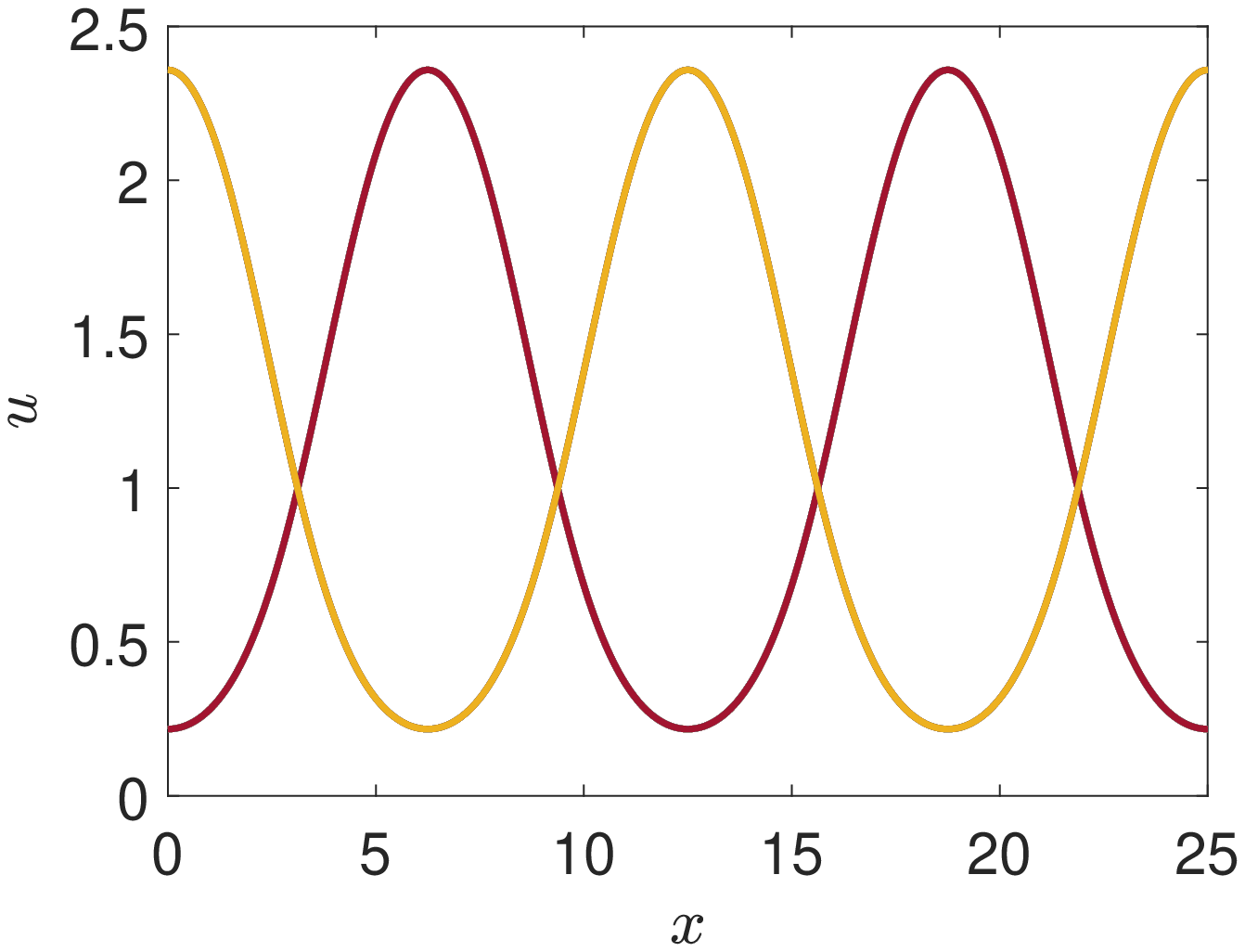}}
    
    \subfloat[Mixed, $R=\min(u_\textrm{Neumann})$]{\includegraphics[width=0.4\textwidth]{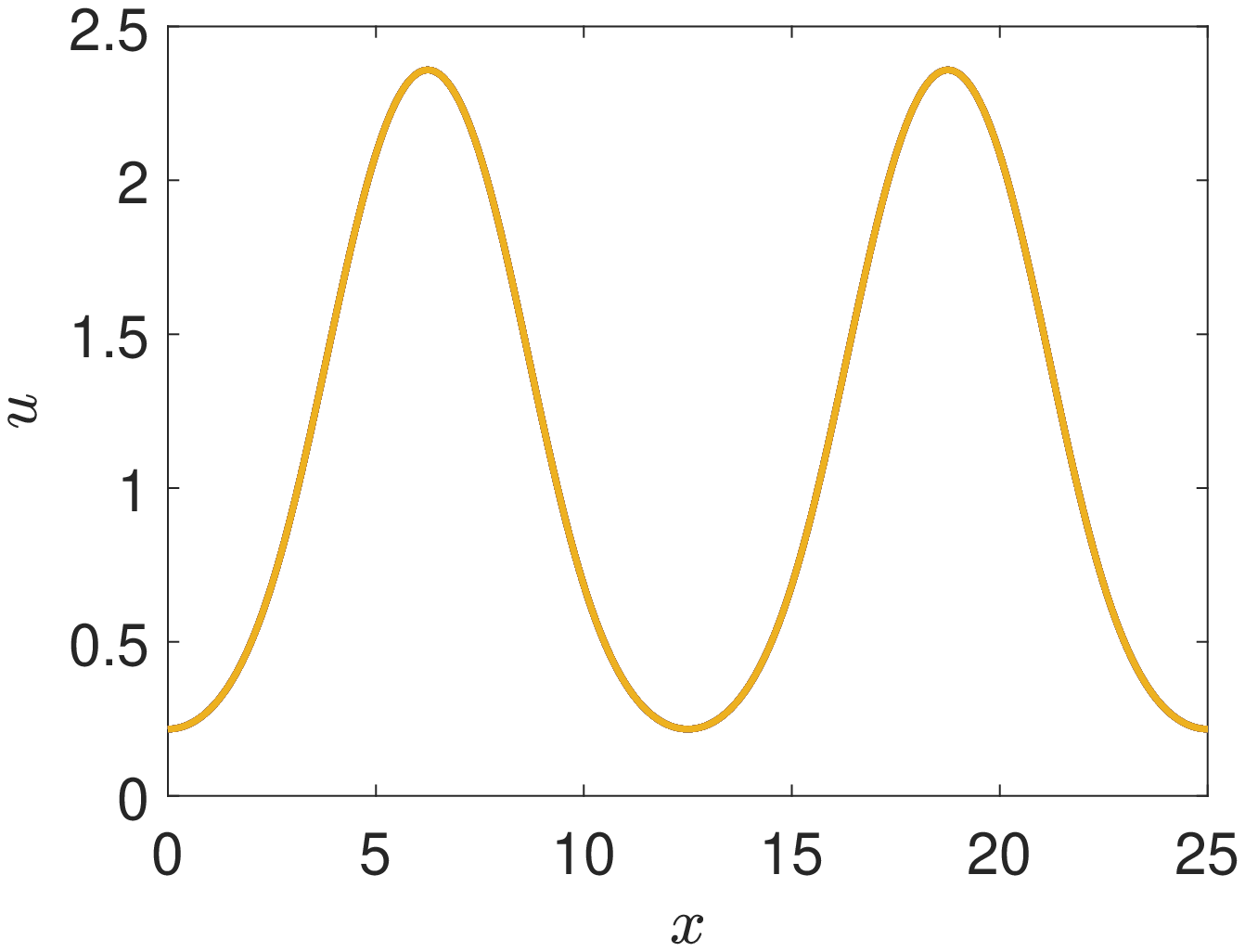}}
    \subfloat[Mixed, $R=0$]{\includegraphics[width=0.4\textwidth]{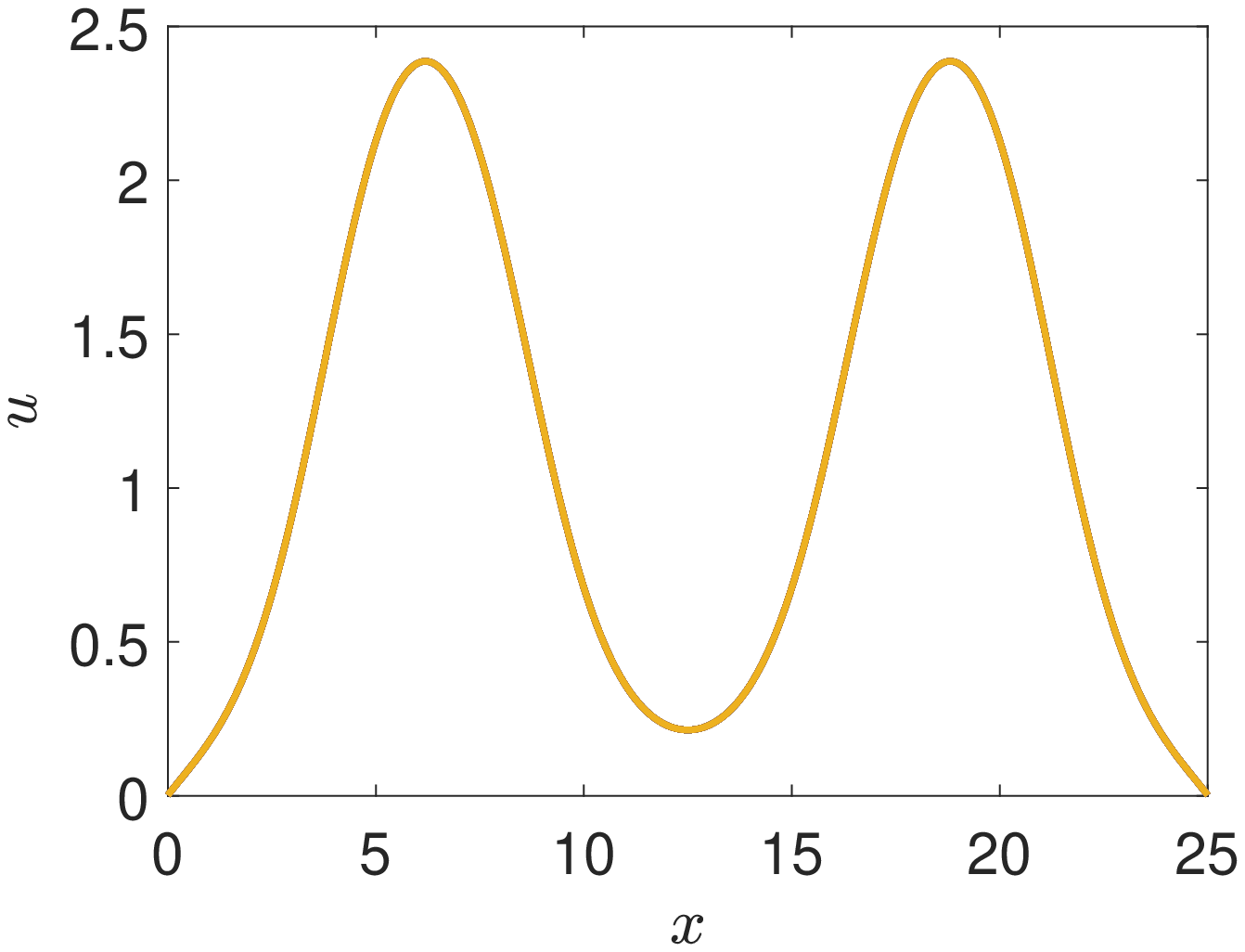}}
    \caption{Steady state solutions from simulations of the Schnakenberg kinetics in Table \ref{table_reactions} under four choices of boundary conditions. Ten simulations are shown where in (a), the initial simulation uses random initial data, but each subsequent simulation is initialized by cyclically shifting this solution $5\%$ along the length $L$, and a stable steady state is found (by cyclic symmetry, these are automatically stable steady states as the first one was). For (b)-(d), the initial data are taken as each of the steady states in (a), evolved for $t=100$ units in time, and then used to find a nearby stable steady state. Note that the Neumann solutions are a subset of the periodic ones. Finally the first set of mixed steady states (c) are, up to numerical accuracy, a subset of the Neumann steady states (b), whereas the second set (a) are close to, but not the same as, these. Note the colours used in (c)-(d) correspond to the last plotted solution, as all ten solution curves fall onto the same points. Parameters used were $a=0.1$, $b=1.7$, $c=1$, $D=20$, and domain length $L=25$. }
    \label{pattern_selection}
\end{figure}

To demonstrate this concretely, we simulate the Schnakenberg kinetics from Table \ref{table_reactions} (though with different domain lengths, $L$) using periodic, Neumann, and mixed boundary conditions. We employ a method-of-lines approach to simulate the RDS, using the standard three-point stencil to discretize the Laplacian. We use the Matlab function \emph{`ode15s'} to evolve these ordinary differential equations in time, and the Matlab function \emph{`fsolve'} to compute steady state solutions to the discretized system on $n=2000$ points. Writing $(u^*,v^*)$ as the homogeneous steady state, we consider an initial condition of the form $(u^*(1+10^{-2}F_i), v^*(1+10^{-2}G_i))$, where $F_i, G_i$ are independent normally distributed random variables with unit variance at each node $i$. Starting with periodic boundary conditions, we evolve this initial condition for $t=10^4$ units of time, and then use its value to find a nearby stable steady state (which is always almost identical to the last transient state, suggesting it is reachable from random initial data). From this steady state solution, we generate nine others, shown in Figure \ref{pattern_selection}(a), by shifting this steady state pattern cyclically. By the translational invariance of the Laplacian with periodic boundary conditions, any stable shifted steady state solution with any shift is still a stable steady state solution, and we confirm this numerically.

We use these ten shifted periodic solutions as initial conditions for simulating the RDS using three other sets of boundary conditions, and then again find stable steady states after evolution in time. We use Neumann conditions in Figure \ref{pattern_selection}(b), mixed conditions given in \eqref{mixed_BCs} with $R$ equal to the minimal value of $u$ from the Neumann case in (c), and conditions \eqref{mixed_BCs} with $R=0$ in (d). We see precisely the pattern selection as described, with the Neumann solutions corresponding to a shifted version of those in the periodic case (hence only admitting two shifted solutions, rather than infinitely many). The mixed conditions with $R$ set as the minimum value of $u$ with Neumann conditions select precisely one solution from the two Neumann solutions (and no others). Finally, using $R=0$ in (d) locally changes the solution from (c) near the boundary, but qualitatively does not change the solution (nor are any new solutions found in this case). Simulations with random initial conditions, as well as repeating this procedure with 100 different periodic solutions as initial data, yield precisely the same curves shown in (b)-(d) as these are the only attracting steady states for all of these different initial data.

\begin{figure}
    \centering
    \subfloat[Periodic]{\includegraphics[width=0.4\textwidth]{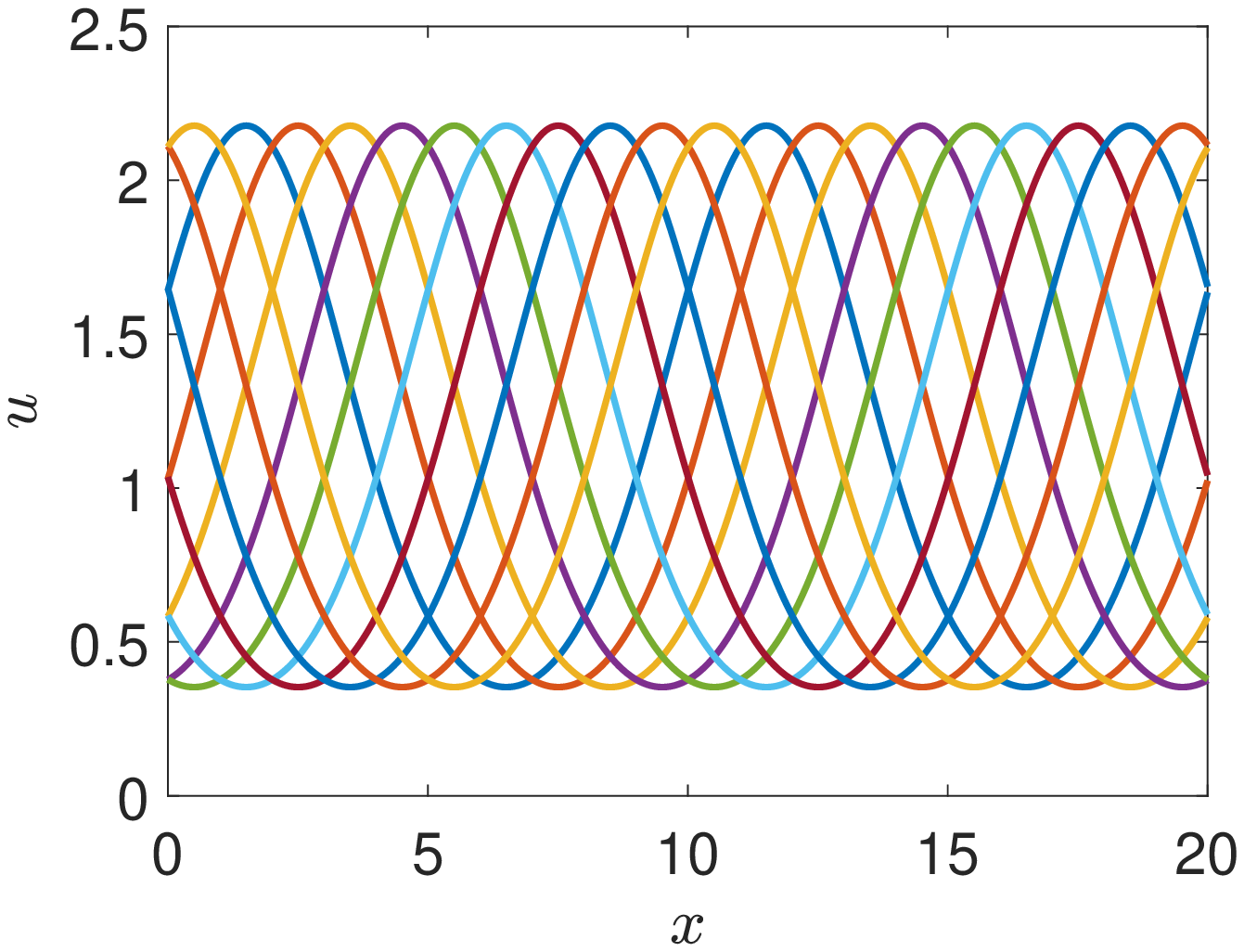}}
    \subfloat[Neumann]{\includegraphics[width=0.4\textwidth]{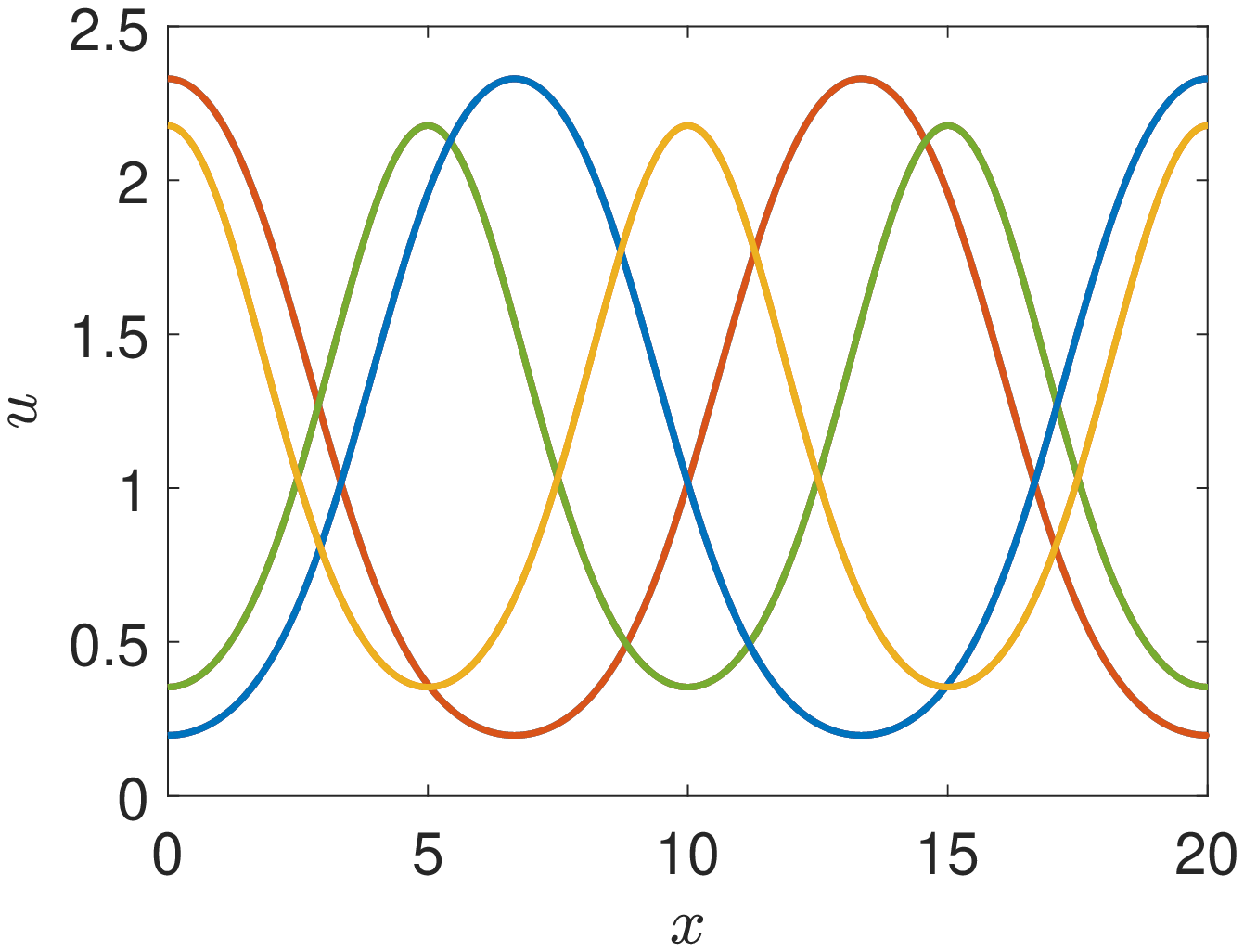}}
    
    \subfloat[Mixed, $R=\min(u_{\textrm{Neumann }\cap \textrm{ periodic} })$  ]{\includegraphics[width=0.4\textwidth]{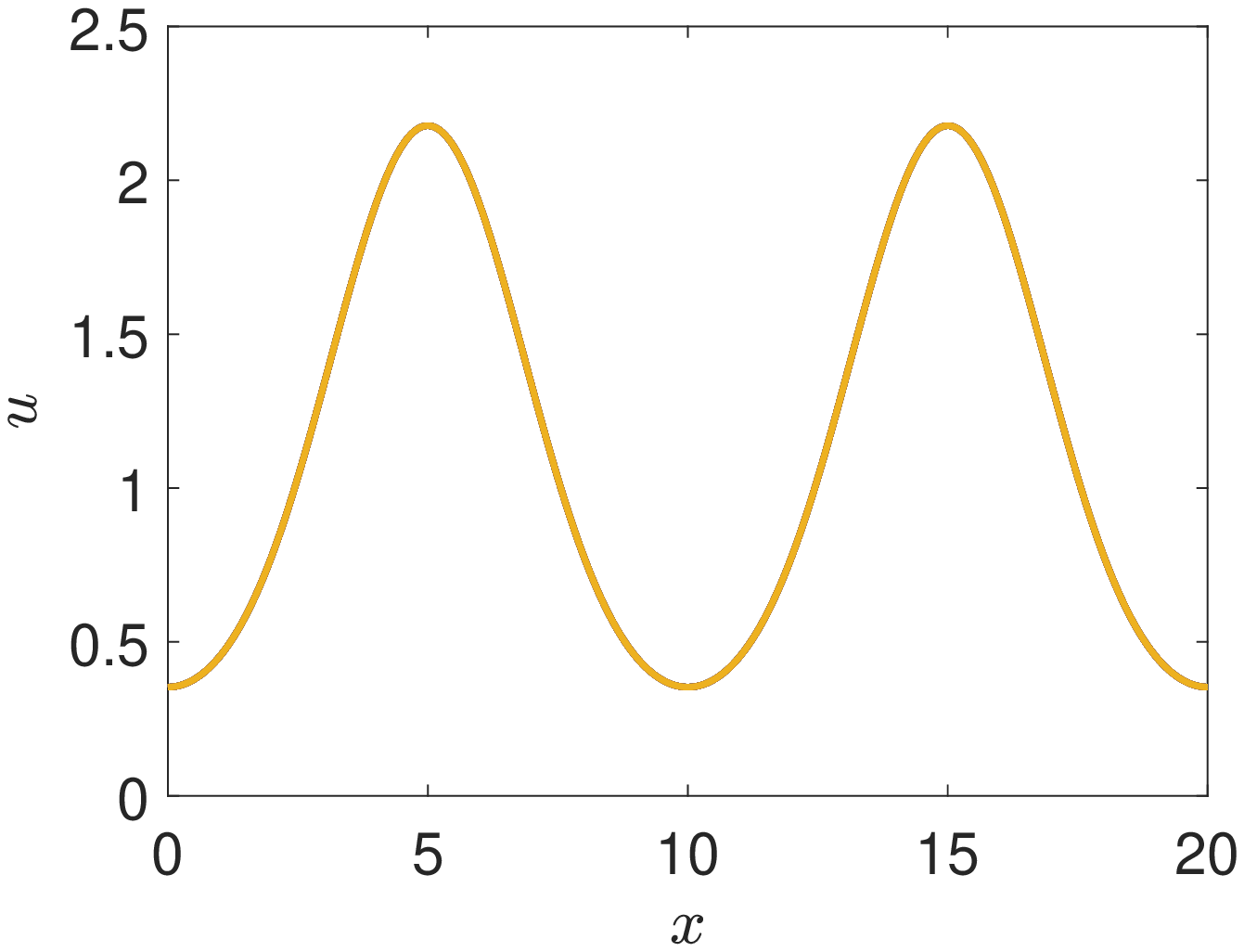}}
    \subfloat[Mixed, $R=0$]{\includegraphics[width=0.4\textwidth]{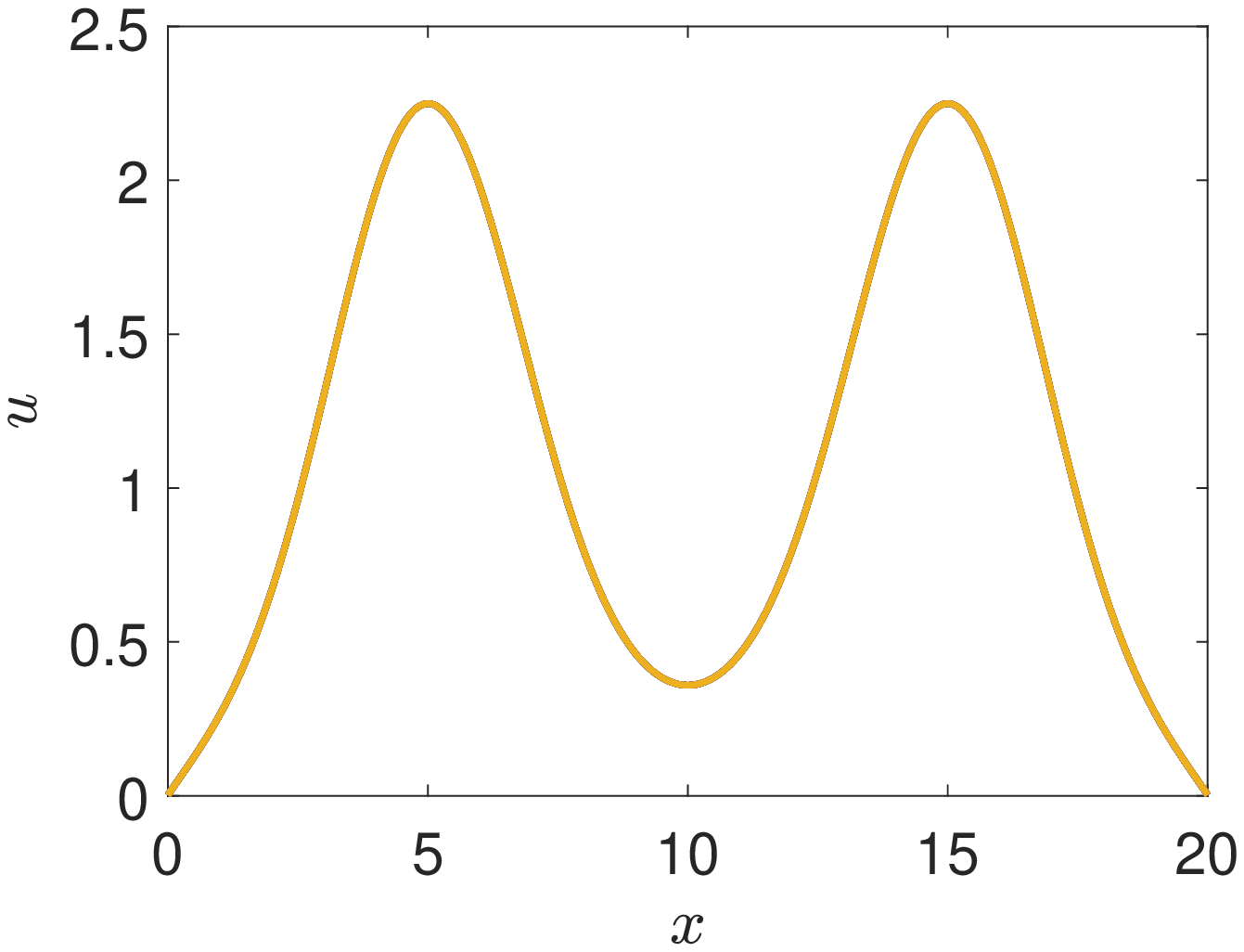}}
    \caption{Steady state solutions from simulations of the Schnakenberg kinetics in Table \ref{table_reactions} under four choices of boundary conditions, exactly as in Figure \ref{pattern_selection} except taking a domain length of $L=20$. Note that in (b), the solutions in red and blue have different amplitudes than those in yellow and green. }
    \label{pattern_selection2}
\end{figure}

More generally, solutions satisfying Neumann boundary conditions need not give rise to periodic patterns, particularly if the domain permits half-integer-modes \citep{murray2004mathematical}. In Figure \ref{pattern_selection2} we give an example of this following the same procedure as in Figure \ref{pattern_selection} with a reduced domain length. Two of the four solutions given in Figure \ref{pattern_selection2}(b) do not correspond to solutions with periodic boundary conditions shown in (a), possessing a different amplitude. Setting $R$ as the minimal value of $u$ from these simulations with Neumann conditions (considering only the solutions which are also periodic, shown in green in (b)), we obtain a single solution with mixed conditions in (c). Setting $R$ equal to either the other minimum of $u$ from the Neumann conditions, or to $0$, has only a small influence on the solution near the boundary, as can be seen in (d). The mixed boundary conditions further exclude half-integer mode solutions.

The reduction of the admissible set of patterns has only been shown numerically for this particular example, and one would need to employ more sophisticated mathematical techniques to demonstrate this for all kinetics and parameters, and beyond one-dimensional examples. Nevertheless, the PDEs in question are local, and so any solution where $v$ is extremal and $u$ is minimal at the same points which satisfy both Neumann and periodic boundary conditions will also satisfy the conditions \eqref{mixed_BCs} when $R$ is suitably chosen (i.e.~exactly or approximately equal to a minimum of a Neumann solution). It is not always the case that reaction-diffusion patterns will have both species sharing extremal points, but this can be shown near a Turing bifurcation via linearization, and is observed (at least approximately) in many systems numerically beyond the bifurcation \citep{dillon1994pattern}. We note that in Figures 3(d) and 4(d), the value of $R$ chosen selects a steady solution with a nonzero boundary flux, suggesting that $u$ is being depleted at the boundaries due to reactions outside of the domain. Hence, this is why we view these as a type of `open' system, which are common in biological systems though less well-studied in general compared to closed systems. The pattern selection mechanism described here also suggests that peaks of the activator $u$ should be approximately half of a wavelength away from the boundary, and we will directly test this prediction across a range of parameters and kinetics in Section \ref{2Dnumerics}.

\section{Demonstrations of Robust Isolated Patterning}\label{numerics_section}

\begin{table}

    \begin{tabular}{| l | c | c | c |c| c | c | c | c | c| c|}
    \hline
    Kinetics  & $f$ &$g$ & $R$ & $L$ &$D$ & $a$ & $b$ & $c$ &$d$ & $e$  \\ \hline
    SCH (Spots) & $a-cu+u^2v$& $b-u^2v$ & 0 & 55 & 1000 & 0.01 & 1.8 & 1 & -& - \\ \hline
    SCH (Labrynthine)  & $a-cu+u^2v$& $b-u^2v$ & 0 & 100 & 20 & 0.01 & 1.7 & 1 & -& -\\ \hline
    GM &$\displaystyle a+\frac{u^2}{v}-bu$ &$u^2-cv$ & 0 & 90 & 1000 & 0.01 & 0.5 & 5.5 &-& -\\ \hline
    TH (Spots) & $\displaystyle a-u-\frac{buv}{1+u+cu^2}$&  $\displaystyle d(e-v)-\frac{buv}{1+u+cu^2}$& 2& 60 & 30 & 92 & 18.5 & 0.1 &1.5& 64 \\\hline
    TH (Labrynthine) & $\displaystyle a-u-\frac{buv}{1+u+cu^2}$&  $\displaystyle d(e-v)-\frac{buv}{1+u+cu^2}$& 2& 60 & 10 & 92 & 18.5 & 0.1 &1.5& 64 \\\hline
    FHN  & $u-au^3+v-b$ &$c-u-dv$ & -1.6 & 250 & 50 & 0.33 & 0.6 & 0.6 &0.99& 1\\
    \hline
    
    \end{tabular}
    \caption{A list of reaction kinetics and non-dimensional parameter regimes simulated, where $L$ is a domain length. SCH: Schnakenberg \citep{schnakenberg1979simple}; GM: Gierer-Meinhardt \citep{gierer1972theory}; TH: Thomas \citep{kernevez1979hysteresis}; FHN: FitzHugh-Nagumo \citep{fitzhugh1955mathematical, fitzhugh1961impulses, nagumo1962active}. Note that the $R$ values chosen here are roughly close to minimal values of typical patterns, but not exactly these values. Note that the FitzHugh-Nagumo kinetics model voltage potential relative to a ground state via the variable $u$, and so need not admit positive solutions of this variable. }\label{table_reactions}
\end{table}

We now give example simulations of RDS with the boundary conditions \eqref{mixed_BCs}, as well as those with Neumann boundary conditions. Additionally, we compare these with simulations of the heterogeneous problem \eqref{ueq_het}-\eqref{veq_het} with Neumann conditions on the outer boundary, to demonstrate that our asymptotic reduction to the mixed boundary conditions is valid for large values of the relaxation rate $\rho$ and suitably thin extended domains. Finally we will also give example simulations in more complicated geometries and higher-dimensional domains, as well as systems with more than two species, showing that the isolated patterning emergent from our mixed boundary conditions extends beyond these initial examples. 

For all of these simulations, we used the commercial finite-element software COMSOL. In the two-dimensional simulations, we used a minimum of $5\times 10^4$ triangular second-order finite elements, and in the three-dimensional simulations we used at least $10^5$ tetrahedral elements. We always chose parameters such that there was a unique homogeneous steady state for the kinetics, given by $f$ and $g$ for homogeneous Neumann boundary conditions\footnote{Note that such an initial condition need not satisfy the boundary conditions \eqref{mixed_BCs}, but will effectively instantly relax to solutions satisfying these conditions at the boundary under time evolution, as the system is parabolic.}, and we perturbed this state by multiplying it by a normally distributed random variable of standard deviation $10^{-2}$ and unit mean, sampled identically and independently at each finite element (as in Section \ref{pattern_selection_sec} in the 1-D case). All simulations shown throughout this section are given at $t=10^5$ units of time, by which point they were within numerical tolerances of a stationary solution. The timestepping was implemented using a standard adaptive backwards-differentiation formula of orders $1$ through $5$. Refinements in space were used to check convergence for particular simulations, and finite-difference simulations were carried out in Matlab (using the five-point Laplacian stencil) to check convergence to the same steady state pattern from identical initial data.

\subsection{Interior Patterning across Two-Species Kinetics}\label{2Dnumerics}

We now describe results from simulations of each of the reaction kinetics and parameter regimes shown in Table \ref{table_reactions}. These are given in Table \ref{table_sim}, where the first column indicates the parameter set, the second column simulations with Neumann boundary conditions, the third column simulations with conditions \eqref{mixed_BCs}, and finally the fourth column shows simulations of the enlarged heterogeneous system \eqref{ueq_het}-\eqref{veq_het}. For each of these simulations we used the same realization of the random initial condition, as all of these systems can admit multistability of different inhomogeneous steady states depending on the initial data \citep{borckmans1995turing, jensen1993subcritical}, and results are displayed for sufficiently long times to ensure transient behaviour has relaxed onto steady states.

Comparing the Neumann simulations with those using our mixed boundary conditions in Table \ref{table_sim}, we can clearly see that the mixed boundary conditions lead to the inhomogeneous patterns being confined to the interior of the domain. Additionally, it is visually apparent in each case that the distance from the peak of an outermost structure (spot or stripe) to the boundary is approximately half of the wavelength between interior structures, despite the fact that $R$ was not set to be exactly the minimum pattern value as in Section \ref{pattern_selection_sec}. Finally, as the same initial perturbations were used across all simulations, comparing the mixed boundary conditions to the heterogeneous ones, we see essentially identical steady state solutions in all cases except for the FitzHugh-Nagumo kinetics, which admit small differences between the steady states selected. We also considered how the asymptotic reduction holds for different domain geometries and for smaller values of $\rho$, but for brevity omit this analysis (largely as it depends heavily on the kinetics and type of pattern studied). Essentially these results confirm our analysis in Section \ref{asymptotic_sec} that the mixed conditions approximate such heterogeneous problems at least once any transients have relaxed. We note that these boundaries have discontinuous derivatives at the corners, but simulations on circular domains have identical properties. Similarly, simulations with much larger external regions (e.g.~when $\tilde{\Omega}$ is $[-L,2L]\times[-L,2L]$, so that the domain length is three times larger) give steady states with interior patterning, as in those on the interior domain with mixed boundary conditions in Table \ref{table_sim}. This suggests that, at least in some cases, the geometric assumptions on the exterior domain can be relaxed, though we will not pursue this further here.

These boundary conditions have some influence on the structure of the resulting patterns beyond the boundaries, especially in the case of the labyrinthine patterns shown in Table \ref{table_sim}. The stripes in the Schnakenberg example show a clear confinement due to the square geometry. In the labyrinthine case for Thomas kinetics, we see that the (roughly) hexagonal arrangement of inverted spots in the Neumann case is forced into a square arrangement in the centre of the domain with the mixed conditions, with a stripe-like structure now surrounding this region. The FitzHugh-Nagumo structures are qualitatively similar on the interior of the domain, with random-looking labyrinthine stripes, but the boundary is clearly marked with a striped region having some local effects on the meandering of the stripes inside. As these labyrinthine solutions are more global in structure than localized spots, we expect less isotropic patterns and more of an influence of the confining geometry and initial conditions. This is true even in the Neumann case, as we can see by the orientation of the Schnakenberg stripes aligning in a large part of the domain with one of the axes. We also remark that the alignment of labyrinthine patterns with Neumann boundary conditions can be sensitive to initial conditions, and one can obtain quantitatively large differences for different realizations of random initial conditions. In contrast, the Schnakenberg and Thomas kinetics with our mixed boundary conditions gave consistently the same final steady state seemingly independent of a variety of initial conditions (though the positioning of the interior structures of the FitzHugh-Nagumo steady states did depend on initial conditions). Depending on the application, robust alignment may or may not be desirable. We will further explore interior patterning and pattern selection effects in the next section by considering more complicated domains.

\begin{table}
\centering
    \begin{tabular}{| Sc | Sc | Sc | Sc|}
    \hline
    Kinetics  &  Neumann BCs  & Mixed BCs & Heterogeneous  \\ \hline
    \rotatebox{90}{\hspace{0.7cm} SCH (Spots)} & \includegraphics[width=0.2\textwidth]{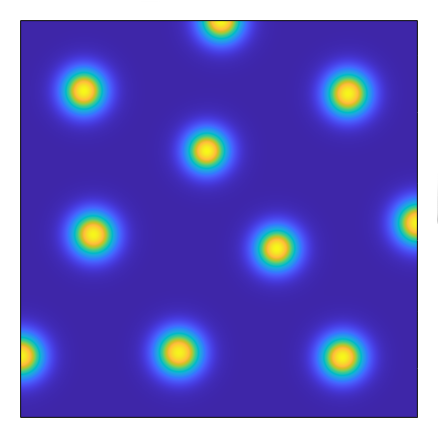} & \includegraphics[width=0.2\textwidth]{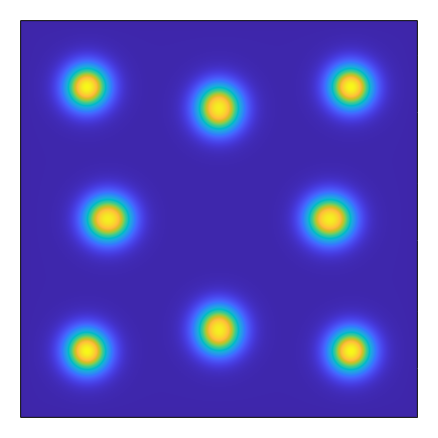} & \includegraphics[width=0.2\textwidth]{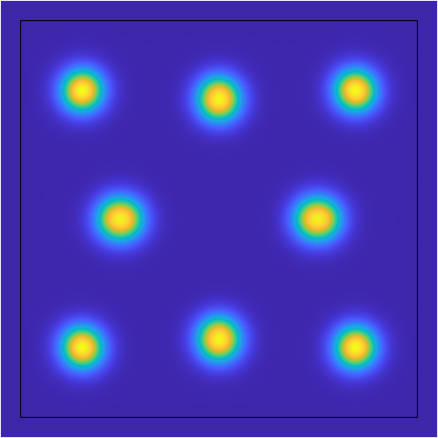}  \\ \hline
    \rotatebox{90}{\hspace{0.35cm} SCH (Labyrinthine)}& \includegraphics[width=0.2\textwidth]{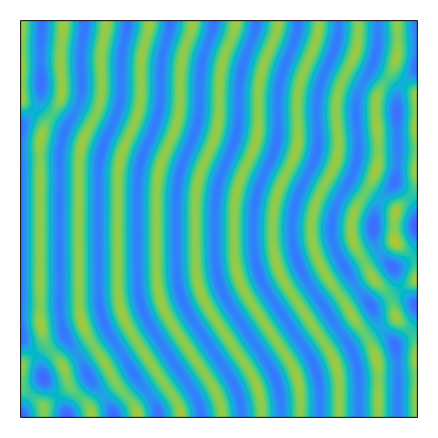} & \includegraphics[width=0.2\textwidth]{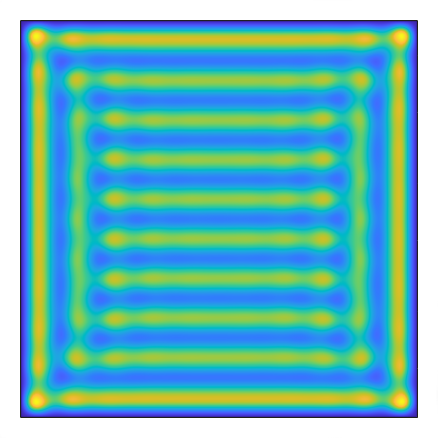} & \includegraphics[width=0.2\textwidth]{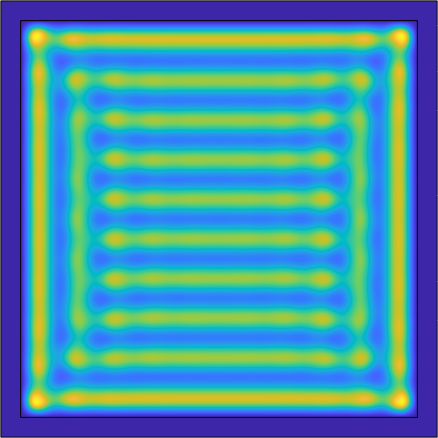}\\ \hline
    \rotatebox{90}{\hspace{1.2cm} GM} & \includegraphics[width=0.2\textwidth]{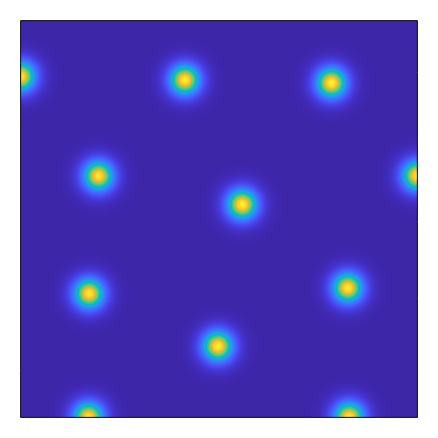} & \includegraphics[width=0.2\textwidth]{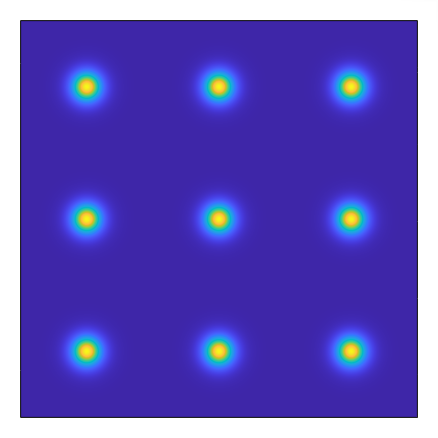} & \includegraphics[width=0.2\textwidth]{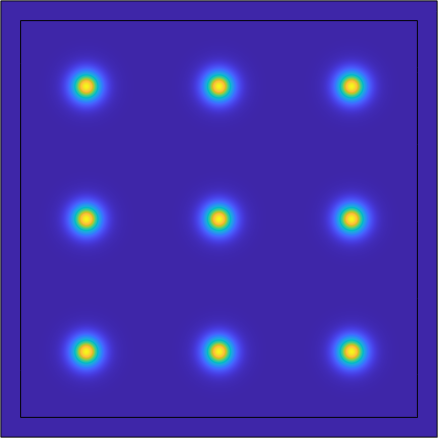}\\ \hline
    \rotatebox{90}{\hspace{0.8cm} TH (Spots)} & \includegraphics[width=0.2\textwidth]{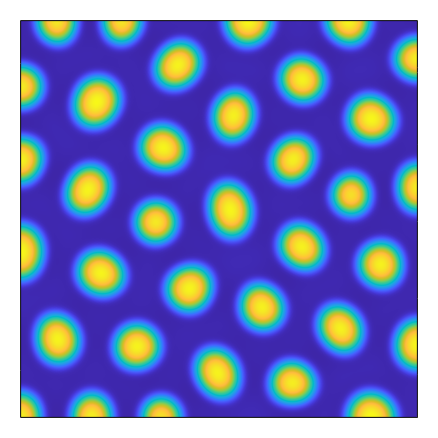} & \includegraphics[width=0.2\textwidth]{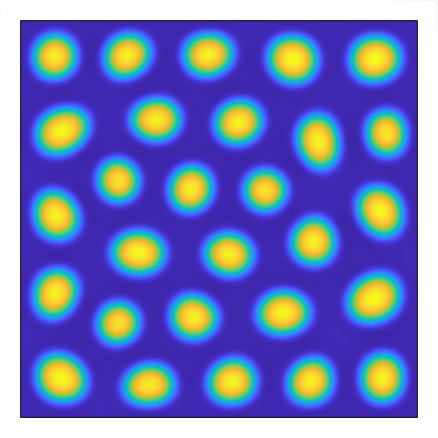} & \includegraphics[width=0.2\textwidth]{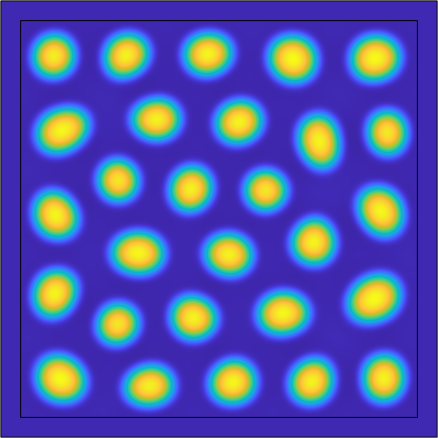}\\ \hline
    \rotatebox{90}{\hspace{0.4cm} TH (Labrynthine)} & \includegraphics[width=0.2\textwidth]{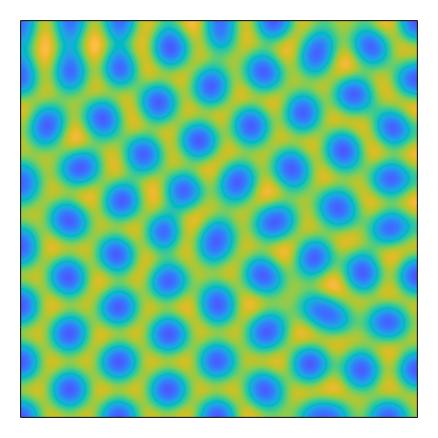}  & \includegraphics[width=0.2\textwidth]{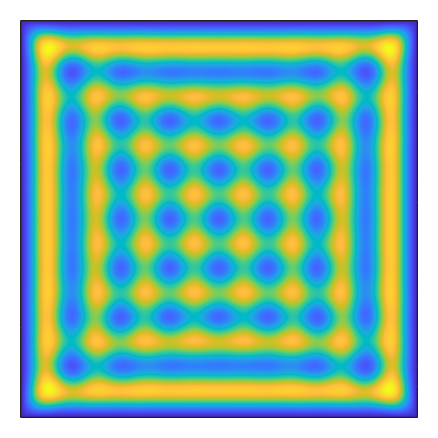} & \includegraphics[width=0.2\textwidth]{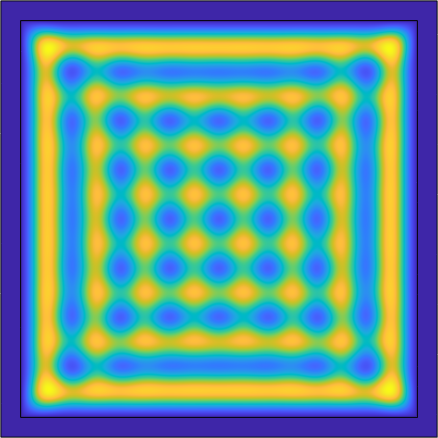} \\ \hline
    \rotatebox{90}{\hspace{1.2cm} FHN} & \includegraphics[width=0.2\textwidth]{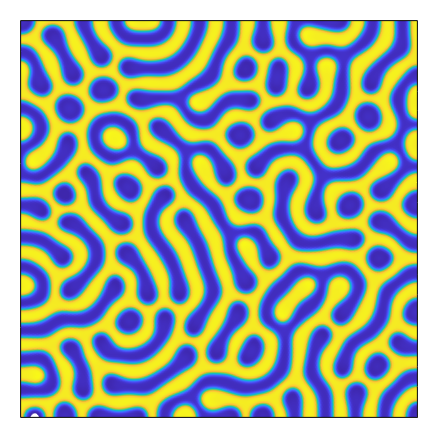}&
    \includegraphics[width=0.2\textwidth]{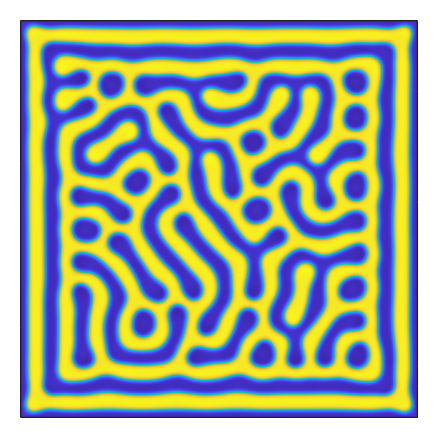}&
    \includegraphics[width=0.2\textwidth]{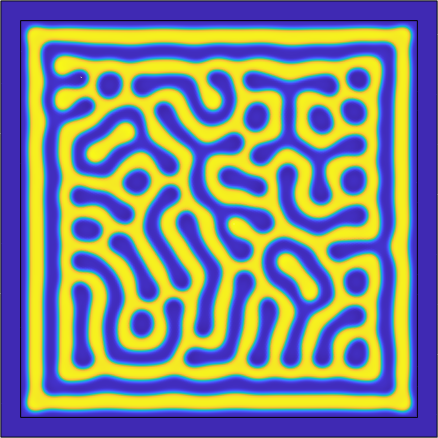}  \\
    \hline
    \end{tabular}
    \caption{Concentrations of $u$ from simulations of the kinetics and parameters given in Table \ref{table_reactions} with Neumann conditions, the mixed boundary conditions \eqref{mixed_BCs}, and in a heterogeneous system given by \eqref{ueq_het}-\eqref{veq_het} with $\rho=10^5$ and Neumann conditions on the exterior boundary. Maximal values of $u$ are in the light yellow regions, and minimal values in the dark blue, though precise numerical values vary between kinetics and parameters, which are specified in full in Table \ref{table_reactions}. The domain in columns two and three was the square $\Omega = [0,L]^2$, and the heterogeneous domain was the enlarged square $\tilde{\Omega} = [-0.05L, 1.05L]^2$, using the same interior domain $\Omega$ as described near Equations \eqref{ueq_het}-\eqref{veq_het}. This final column has an interior domain $\Omega$ corresponding to the black square, where the boundaries in the second and third columns are located. For each set of kinetics, the colour scale is fixed across the row, and the approximate minimum/maximum (blue/yellow) values are as follows: SCH (Spots) 0/23.56; SCH (Labyrinthine) 0/3.58; GM 0/163.9; TH (Spots) 1.56/37.4; TH (Stripes) 2/16.2; FHN -1.65/1.65}\label{table_sim}
\end{table}

We remark that the solutions obtained are reasonably robust to choices of $R$, which were taken to be close to the minimal value of $u$ in the Neumann solutions. Specifically, variations of $0.1$ in any direction had no effect except in locally changing the value of $u$ near the boundary, and larger variations in $R$ led to local changes with some impact on pattern selection, but no real qualitative influence on the structure of solutions away from the boundary. The lengthscales $L$ were chosen to be sufficiently large to avoid finite-size effects of a few localized structures. Finally we remark that the lengthscale of the outer domain had no influence on the resulting patterns when it was increased to a width of $0.5L$, suggesting a robustness to the qualitative effect extending beyond the thin-domain assumption made in Section \ref{asymptotic_sec}. As this assumption is not needed in the 1-D case, we suspect it is a technicality not affecting the generic situation and could in principle be relaxed, though we do not pursue this here.

\subsection{Higher-Dimensions \& Complex Geometries}\label{geometry_sec}

We now demonstrate the effects of more complicated domain geometries on patterns with the mixed boundary conditions \eqref{mixed_BCs}. As seen in the simulations in the previous section, both spot and labyrinthine patterns were more ordered with these conditions, compared to the Neumann case. In particular, the labyrinthine patterns shown in Table \ref{table_sim} conform to the geometry in the case of the mixed conditions \eqref{mixed_BCs}, whereas the Neumann patterns were seemingly less affected by the boundary, having a broadly more isotropic character. We now show how these labyrinthine patterns are influenced by more complicated domain boundaries with Neumann and our mixed conditions. We then demonstrate that interior confinement, as well as these pattern selection effects, extend to higher spatial dimensions.

Throughout these examples, we will use the Schnakenberg kinetics from Table \ref{table_reactions}, and primarily focus on parameter regimes which give rise to labyrinthine solutions, comparing qualitative features such as stripe orientation and defects.  We remark that for periodic boundary conditions and small-amplitude patterns near the onset of a Turing instability, there is a dichotomy between stripe and spot solutions \cite{ermentrout1991}, determined by signs and relative magnitudes of quadratic and cubic interactions. More recent work has shown that such a simple classification does not hold beyond the weakly nonlinear regime, and qualitatively different kinds of patterns can be simultaneously stable or interact \citep{borckmans1995turing, jensen1993subcritical, Bozzini-2015-WNA}. We observe here that the largest qualitative differences are determined by the geometry (and the initial data in the Neumann setting) in the labyrinthine cases. We focus on defects (regions where the stripes are no longer contiguous at the highest value of $u$, which appear like dappled spots), in addition to stripe orientation.

\begin{figure}
    \centering
    \subfloat[Neumann]{\includegraphics[height=0.15\textwidth]{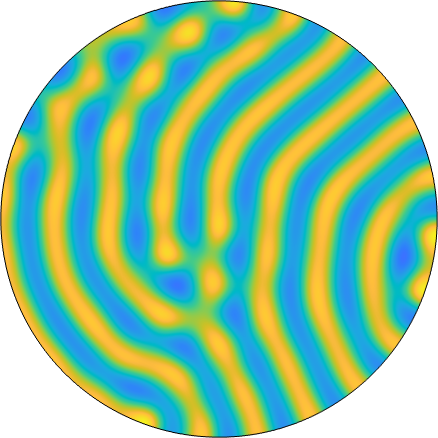}} \hspace{0.2cm} \hspace{0.2cm}\subfloat[Neumann]{\includegraphics[height=0.15\textwidth]{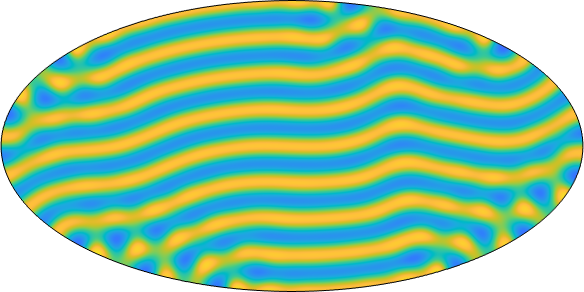}} \hspace{0.2cm}
    \subfloat[Neumann]{\includegraphics[height=0.15\textwidth]{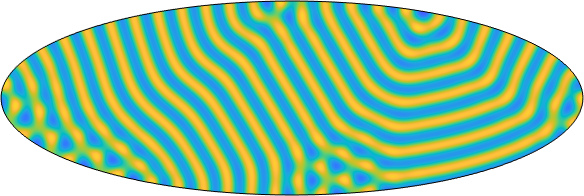}}

    \subfloat[Mixed]{\includegraphics[height=0.15\textwidth]{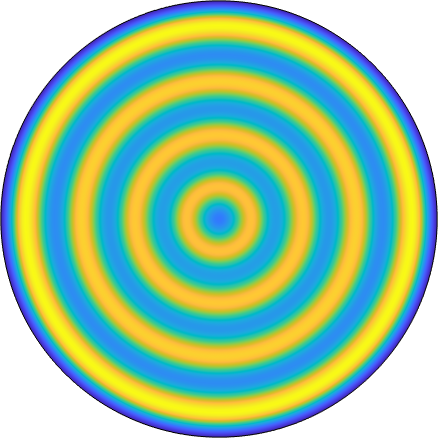}} \hspace{0.2cm} \hspace{0.2cm}\subfloat[Mixed]{\includegraphics[height=0.15\textwidth]{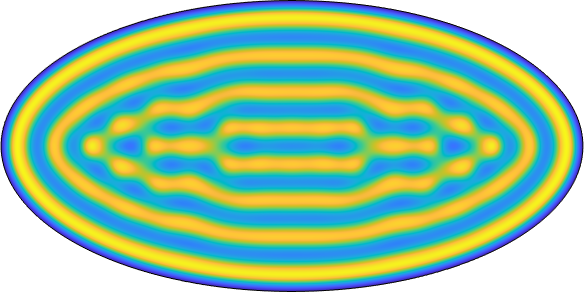}} \hspace{0.2cm}
    \subfloat[Mixed]{\includegraphics[height=0.15\textwidth]{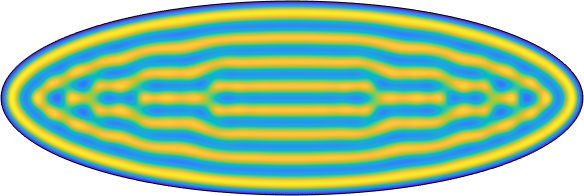}}

    \caption{Values of $u$ computed with Neumann and the mixed boundary conditions \eqref{mixed_BCs} on the interior of circles and ellipses. Simulations shown are on a domain with a semi-minor axis of $40$, and semi-major axes of $40$ in (a) and (d), $80$ in (b) and (e), and $120$ in (c) and (f). We used the Schnakenberg kinetics in Table \ref{table_reactions} with $a=0.01$, $b=1.7$, $c=1$, $D=20$, and $R=0$ in the mixed case. Colour scales were fixed between a minimum (blue) of $0$ and a maximum (yellow) of $2.87$.}
    \label{ellipse_fig}
\end{figure}

First we consider circular and elliptical domains in Figure \ref{ellipse_fig}. As anticipated, solutions with Neumann boundary conditions in Figure \ref{ellipse_fig}(a)-(c) exhibit orientations which are seemingly independent of the domain geometry, though there is some correlation of stripe orientation. Stripe defects appear in all cases with Neumann boundary conditions, mostly near the boundaries where stripes of different orientation intersect. In comparison, there is clearly more symmetry in the confined patterns generated by mixed boundary conditions in Figure \ref{ellipse_fig}(d)-(f), which broadly maintain orientations consistent with the domain geometry. Defects with mixed boundary conditions only occur in the elliptical cases, and seem to be where the stripe curvature is greatest. 

In each of the cases with Neumann boundary conditions, there are slightly different numbers of stripes (depending on what one classifies as a stripe), with (c) having the largest number for these simulations. Different random initial conditions can lead to substantial variation in the number of such structures, depending on what orientation most stripes take. In contrast, if one considers a contiguous region of higher-than-average values of $u$ as a stripe, then the confined patterns generated by mixed boundary conditions all have exactly four of these. This is despite the fact that the largest ellipse occupies three times the area of the circle, and these observations appear to be robust to different random perturbations of the initial conditions. Finally, these results are relatively insensitive to the value of $R$, as simulations with $R=0.2$ and $R=0.7$ (not shown) gave rise to qualitatively identical patterns, with the only observable difference being the minimal value taken. These values were chosen as the minimum of $u$ in the Neumann cases was $u \approx 0.62$.

Next we consider more complicated domains by looking at those formed by perturbing the circle with periodic polar functions. We consider the following parametric domain boundary:
\begin{equation}\label{squiggly_param}
x(s) = L\cos(s)\sqrt{1+\gamma\sin(6s)}, \quad y(s) = L\sin(s)\sqrt{1+\gamma\sin(6s)}, \quad \textrm{for } s\in[0,2\pi],
\end{equation}
where $L$ is now a measure of the domain length (the radius of the circle for $\gamma=0$), and $\gamma \in [0,1)$ signifies the deviation from this circle. The $\sin(6s)$ term makes the perturbations resemble a 6-lobed polar rose or rhodonea curve.

\begin{figure}
    \centering
    \subfloat[Neumann, $\gamma=0.2$]{\includegraphics[width=0.3\textwidth]{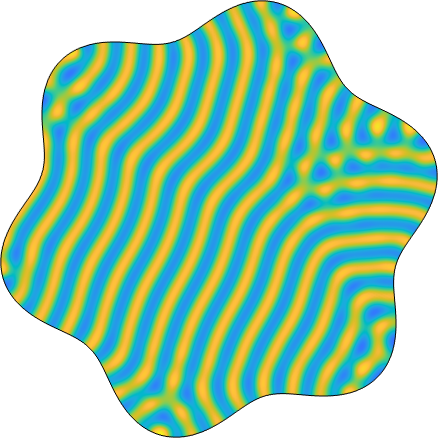}} \hspace{0.2cm}\subfloat[Neumann, $\gamma=0.5$]{\includegraphics[width=0.3\textwidth]{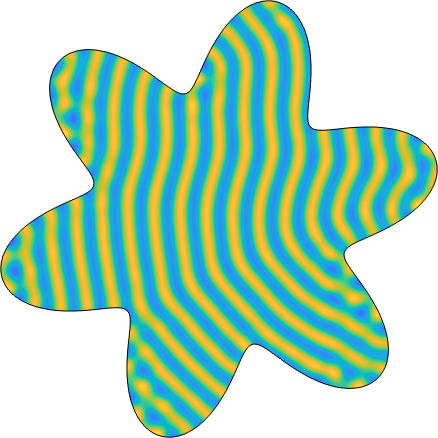}}
    \subfloat[Neumann, $\gamma=0.8$]{\includegraphics[width=0.3\textwidth]{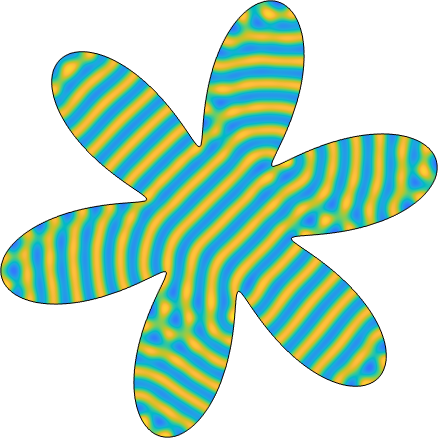}}

    \subfloat[Mixed, $\gamma=0.2$]{\includegraphics[width=0.3\textwidth]{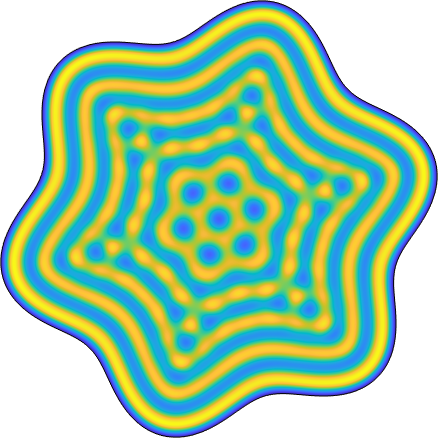}} \hspace{0.2cm}\subfloat[Mixed, $\gamma=0.5$]{\includegraphics[width=0.3\textwidth]{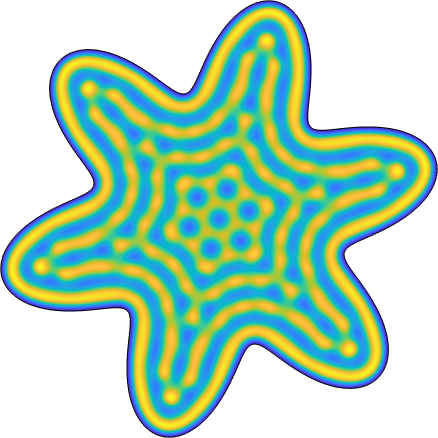}}
    \subfloat[Mixed, $\gamma=0.8$]{\includegraphics[width=0.3\textwidth]{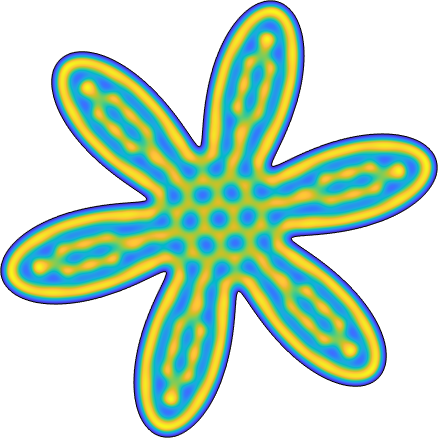}}

    \caption{Values of $u$ computed with Neumann boundary conditions, as well as the mixed boundary conditions \eqref{mixed_BCs} on the interior of the domain given parametrically in \eqref{squiggly_param}. Simulations shown are on a domain with size characterized by $L=70$ using the Schnakenberg kinetics in Table \ref{table_reactions} with $a=0.01$, $b=1.7$, $c=1$, $D=20$, and, in the mixed case, $R=0$. Colour scales were fixed between a minimum (blue) of $0$ and a maximum (yellow) of $2.91$. }
    \label{squiggle_fig}
\end{figure}

In Figure \ref{squiggle_fig} we demonstrate patterns on such domains for three different values of $\gamma$. As in the elliptical case, the solutions with only Neumann conditions in (a)-(c) have stripes with similar alignments, and defects when they meet stripes with different orientations. Very little direct effect of the boundaries can be seen in these simulations, with the patterns broadly consisting of seemingly randomly oriented stripes (with orientations depending on initial data). In contrast, the confined patterns in Figure \ref{squiggle_fig} clearly conform to the boundary, with defects again appearing at regions where the stripes (but not necessarily the boundaries) might be expected to have large curvature. As before, while the structure of the Neumann solutions can vary tremendously depending on the initial data, the mixed boundary conditions lead to consistent patterns even when the initial data are different. In the centre of each of these domains, independent of the value of $\gamma$, is a pattern of seven inverted spots (one surrounded by six others). These simulations demonstrate an intriguing possibility of selecting for hybrid stripe and spot solutions in the domain through the discrete symmetry of the geometry. As in the elliptical case, qualitatively similar patterns are seen for $R=0.2$ and $R=0.7$.

We now consider the same comparisons for three-dimensional patterns. Classification of patterns in higher dimensions is even more difficult than in the two-dimensional case, though there has been work on specific model systems \citep{de1992three, de1997twist, shoji2007most, leda2009instabilities}. As the number of Turing-type structures that can exist in three dimensions is vast, we only give two examples to demonstrate the influence of these boundary conditions. To visualize these patterns, we will only plot the solution on the tetrahedral finite elements above some threshold, so that the observed structures correspond to regions of high concentration of $u$.

\begin{figure}
    \centering
    \subfloat[Neumann]{\includegraphics[width=0.4\textwidth]{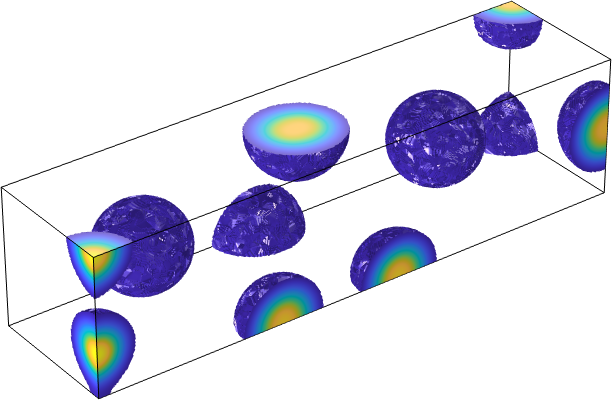}} \hspace{0.2cm}\subfloat[Mixed]{\includegraphics[width=0.4\textwidth]{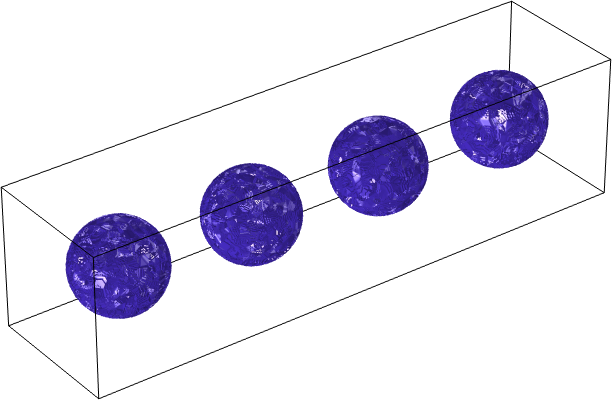}}
    \caption{Thresholded values of $u$ from simulations of \eqref{ueq}-\eqref{veq} using purely Neumann boundary conditions in (a), and the mixed conditions \eqref{mixed_BCs} in (b). Triangular elements were only shown for values of $u>3$ in order to visualize these three-dimensional patterns. Simulations shown are on a domain of size $15 \times 15 \times 60$ using the Schnakenberg kinetics in Table \ref{table_reactions} with $a=0.1$, $b=1.8$, $c=1$, $D=10^3$, and, in the mixed case, $R=0$. The maximum value of $u$ (yellow) is $23.56$, and the minimum (blue) is $0$. }
    \label{3Dspotsfig}
\end{figure}

\begin{figure}
    \centering
    \subfloat[Neumann]{\includegraphics[width=0.35\textwidth]{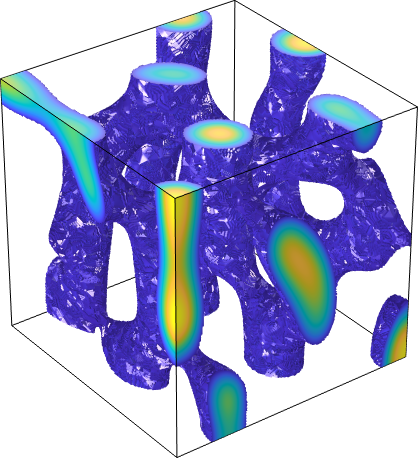}} \hspace{0.2cm}\subfloat[Neumann]{\includegraphics[width=0.35\textwidth]{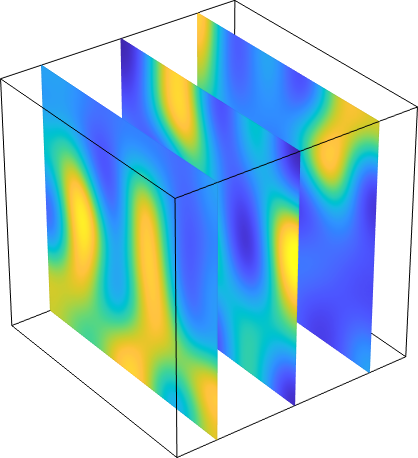}}
    
    \subfloat[Mixed]{\includegraphics[width=0.35\textwidth]{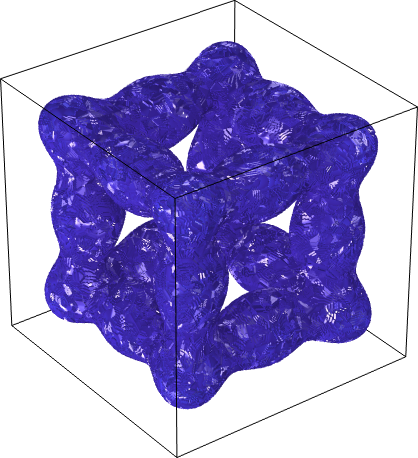}} \hspace{0.2cm}\subfloat[Mixed]{\includegraphics[width=0.35\textwidth]{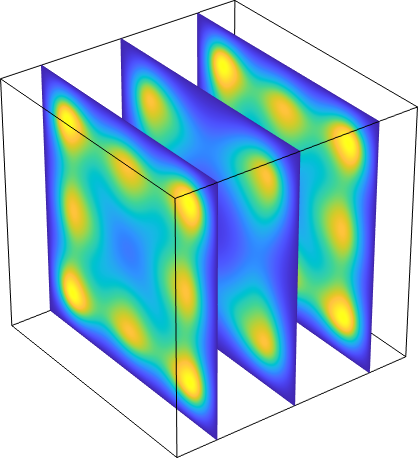}}
    \caption{Thresholded values of $u$ from simulations of \eqref{ueq}-\eqref{veq} using Neumann boundary conditions in (a) (with slices shown in (b)), and the mixed conditions \eqref{mixed_BCs} in (c) (with slices shown in (d)). For the plots in (a) and (c), triangular elements were only shown for values of $u>2.5$ in order to visualize these three-dimensional patterns. Simulations shown are on a domain of size $25 \times 25 \times 25$ using the Schnakenberg kinetics in Table \ref{table_reactions} with $a=0.1$, $b=1.7$, $c=1$, $D=30$, and, in the mixed case, $R=0$. The maximum value of $u$ (yellow) is $3.58$, and the minimum (blue) is $0$.}
    \label{3Dstripesfig}
\end{figure}

In Figure \ref{3Dspotsfig}, we give examples of localized sphere-like structures which emerge again in the Schnakenberg kinetics in a rectangular domain. As expected from the two-dimensional examples, spheres in the Neumann case form hemispheres and quadrants (quarter-spheres), along the boundary. In contrast, the mixed boundary conditions lead to symmetrical spheres of the solution $u$ along the centre of the domain, approximately half of a wavelength away from the boundary in any direction as in the two-dimensional case. As in the case of spots in the two-dimensional examples given in Table \ref{table_sim}, it is clear that the patterns with these boundary conditions are substantially more ordered and symmetrical. Additionally, we find a robustness of pattern structure across different random initial conditions which contrasts with the case of Neumann boundary conditions.

In Figure \ref{3Dstripesfig}, we give an example where more lamella-like structures (analogous to stripes in two-dimensions) appear. These patterns are more complicated and harder to visualize, so cross-sections have been included to help picture them beyond the regions of high activator concentration. The Neumann case admits complicated tube-like structures inside the domain, as well as forming partial tubes along the boundary. In contrast, the confined patterns resemble a face-centred cubic lattice with tubular connections between sphere-like regions of high activator concentration. This example clearly shows a difference between the seemingly more random appearance of patterns in the Neumann case, compared to the extremely structured confined patterns arising from the mixed boundary conditions. 

\begin{figure}
    \centering
    \subfloat[TH (Spots)]{\includegraphics[width=0.35\textwidth]{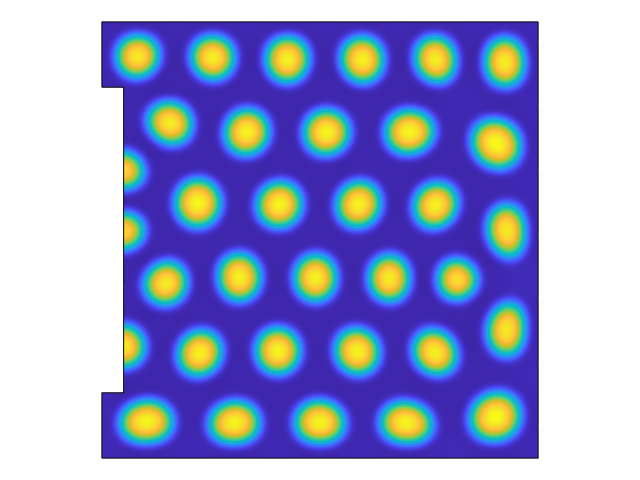}} \hspace{0.2cm}\subfloat[TH (spots)]{\includegraphics[width=0.35\textwidth]{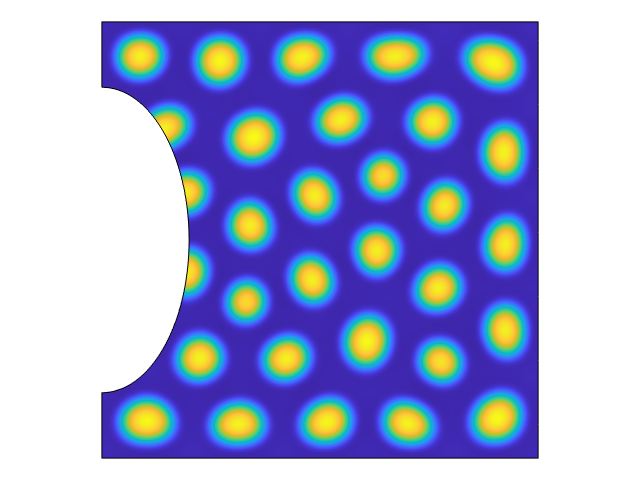}}
    
    \subfloat[TH (stripes)]{\includegraphics[width=0.35\textwidth]{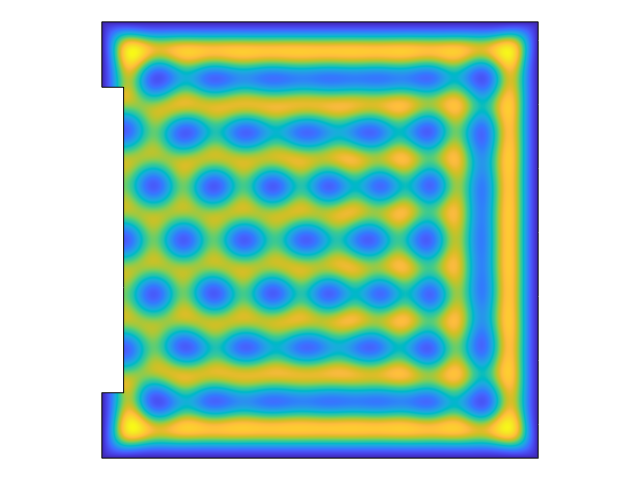}} \hspace{0.2cm}\subfloat[TH (Stripes)]{\includegraphics[width=0.35\textwidth]{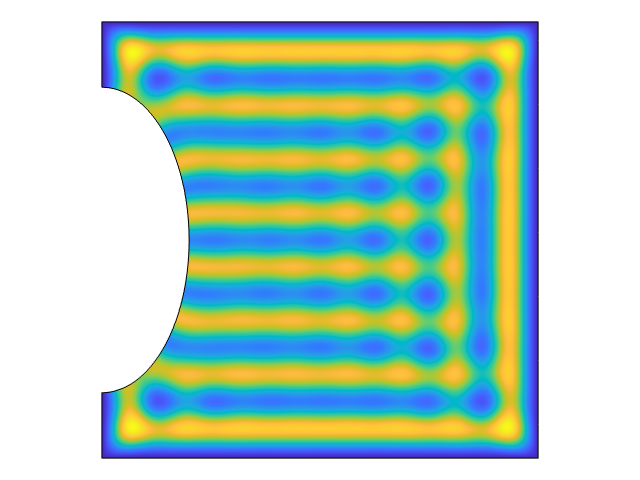}}
    \caption{Values of $u$ from simulations of \eqref{ueq}-\eqref{veq}. The domain is the same as in Table \ref{table_sim} except that a region of the domain has been removed from the left boundary. In (a) and (c), a rectangular region of height $0.7L$ and width $0.05L$ has been removed from the domain, and in (b) and (d) an elliptical region of semi-major axis of $0.35L$ and semi-minor axis $0.2L$ has been removed. We use the mixed conditions \eqref{mixed_BCs} along all of the square boundaries, and Neumann conditions along the three rectangular cut boundaries in (a) and (c), and the elliptical cut boundaries in (b) and (d).  Simulations shown are on a domain of size $L=60$ using the Thomas kinetics and parameters from Table \ref{table_reactions} as labelled, with maximal and minimal values of $u$ as in Table \ref{table_sim}.}
    \label{cutfig}
\end{figure}

Lastly in this section, we consider domains which are composed of both Neumann and the mixed boundary conditions \eqref{mixed_BCs}. Such a setting may arise from cutting a piece of tissue from a patterning field and then inserting an impermeable material, with local no-flux boundaries along the cut. In Figure \ref{cutfig}, we give examples of spots and stripes in the Thomas kinetics from Table \ref{table_sim}, where we have shown rectangular and elliptical `cuts' in the domain along the left boundary, so that these boundaries have Neumann conditions but all others have the mixed conditions.  We see half-spots forming only along these Neumann boundaries in panels (a) and (b). In panels (c) and (d) we observe a more pronounced selection effect of the Neumann boundary conditions, where the domain with the small rectangle removed has an internal structure consistent with the corresponding simulation from Table \ref{table_sim} (though with patterns forming up to the Neumann boundary), but the elliptical cut induces a selection of horizontal stripe patterns even away from this boundary. Such a selection mechanism could be a useful approach to validating these boundary conditions in experimental systems.

\subsection{Extensions to Many-Species}\label{many_species}

Interior pattern localization from mixed boundary conditions can be extended beyond two-species systems. Pattern formation in three and more species systems is far richer than in the two-species case, even when restricting to patterns arising from Turing instabilities \cite{pearson1992pattern, satnoianu2000turing}, and such multispecies kinetics have been studied extensively in recent years in developmental contexts \cite{klika2012influence, diego2018key, scholes2019comprehensive}. We will give a brief example, motivated by pattern formation in three-species Lotka-Volterra systems. We note that two-species Lotka-Volterra systems do not admit patterns except in the case of convex domains and bistability \cite{kishimoto1985spatial, kurowski2017two}, though three-species models do \cite{taylor_patterns_2019}.

We consider the system,
\begin{equation}\label{threespecs_LV}
    \dt{u_i} = D_i \nabla u_i + u_i\left (1-\sum_{j=1}^3 a_{ij}u_j \right ), \quad \textrm{for }i=1,2,3,
\end{equation}
where $D_i>0$ are the diffusion coefficients of each species and $a_{ij} \in \mathbb{R}$ are the interaction coefficients. This is a particular non-dimensionalization of a standard generalised Lotka-Volterra model with diffusion modelling random dispersal. The form of interactions between species permits competition for resources, as well as predation from a generalist or intraguild predator who is also competing for resources with the prey; see \cite{taylor_patterns_2019} for more details about the ecological interpretation. We will use the following extension of the two-species mixed boundary conditions \eqref{mixed_BCs},
\begin{equation}\label{threespecs_mixed_BCs}
    u_1(\bm{x},t)=R, \quad \bm{n}\cdot \nabla u_2=\bm{n}\cdot \nabla u_3=0, \quad \textrm{for all } \bm{x} \in \partial \Omega.
\end{equation}
We will also consider homogeneous Neumann conditions (on all species) for comparison, again simply referring to these as Neumann conditions. 

\begin{figure}
    \centering
    \subfloat[Neumann]{\includegraphics[width=0.23\textwidth]{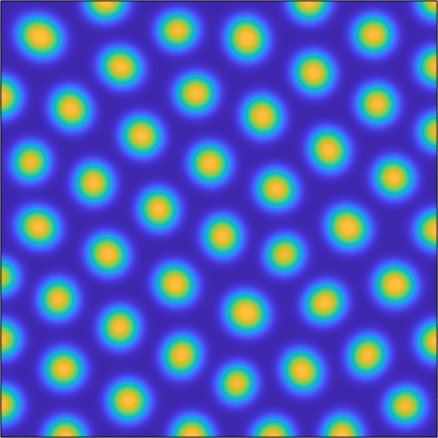}} \hspace{0.2cm}\subfloat[Mixed]{\includegraphics[width=0.23\textwidth]{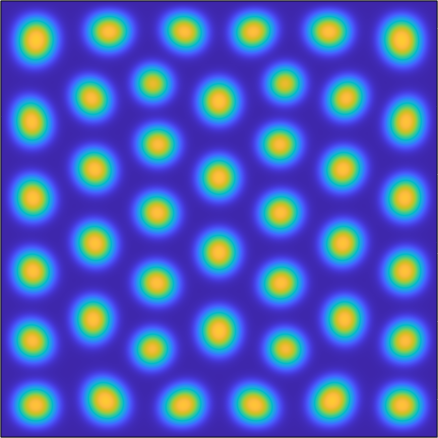}}
    \hspace{0.2cm}\subfloat[Neumann]{\includegraphics[width=0.23\textwidth]{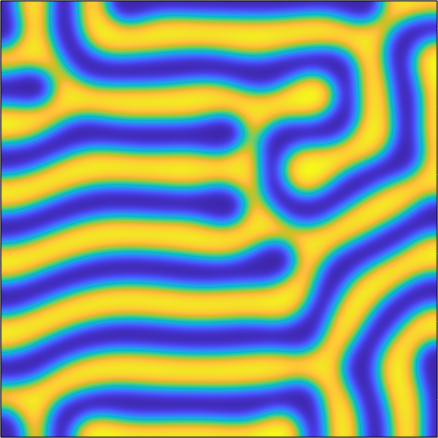}}\hspace{0.2cm}\subfloat[Mixed]{\includegraphics[width=0.23\textwidth]{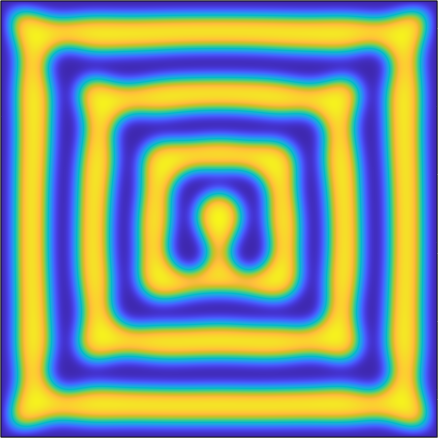}}
    
    \caption{Values of $u_1$ from simulations of \eqref{threespecs_LV} with either Neumann or the mixed boundary conditions \eqref{threespecs_mixed_BCs}. We used $a_{11}=1.1$, $a_{12}=a_{13}=a_{21}=0.8$, $a_{23}=0.8$, $a_{31}=1.7$, $a_{32}=-1$, $a_{33}=1.1$, $D_1=0.1$, $D_2=100$, and $D_3=1$, with $a_{22}=1$ in (a)-(b) and $a_{22}=1.3$ in (c)-(d). $\Omega$ was taken to be a square domain of side length $L=300$. $u_2$ is in phase with $u_1$, and $u_3$ is out of phase with these. The colours correspond to a maximal value of $0.7$ (light yellow) and a minimal value of $0$ (dark blue). Simulations are shown at time $t=10^5$, and were initialized using normal perturbations of the homogeneous steady state as described in the beginning of the section. See Figure 1 of \cite{taylor_patterns_2019} for an ecological description and interpretation of the parameters. We took $R=0$ in the case of mixed boundary conditions. The colour scale ranges between $0$ (blue) and $0.7$ (yellow) in all panels. }
    \label{threespecs_fig}
\end{figure}

We show simulations of this system with parameters giving spots and labyrinthine patterns in Figure \ref{threespecs_fig} for both Neumann boundary conditions, and the mixed boundary conditions \eqref{threespecs_mixed_BCs}. As in the two species simulations above, we see the mixed boundary conditions forcing the patterned regions of $u_1$ into the interior of the domain, and some pattern modulation due to nonlinear interactions between patterned regions. For instance, in (b) we see a few smaller spots due to crowding, and in (d) we see a clear mode selection from the boundaries as shown before in the labyrinthine examples. Of course both of these kinds of patterns can also occur with Neumann conditions, but may be more prevalent for mixed boundary conditions as they confine patterns to the interior. Besides these influences, the patterns are qualitatively similar on the interior of the domain, especially between (a) and (b). Simulations in an extended domain, analogous to the system \eqref{ueq_het}-\eqref{veq_het}, gave the same results as using \eqref{threespecs_mixed_BCs} (not shown).

\begin{figure}
    \centering
    \subfloat[Neumann]{\includegraphics[width=0.23\textwidth]{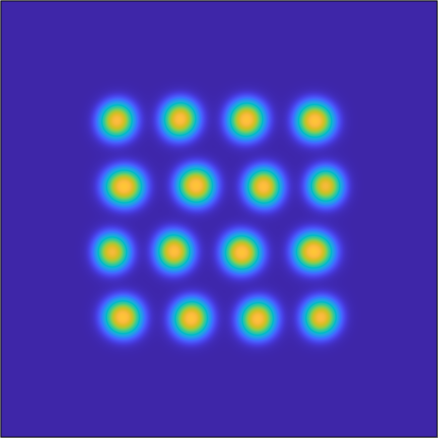}} \hspace{0.2cm}\subfloat[Mixed]{\includegraphics[width=0.23\textwidth]{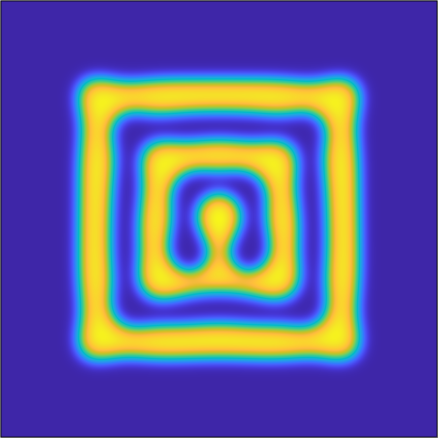}}
    \hspace{0.2cm}\subfloat[Neumann]{\includegraphics[width=0.23\textwidth]{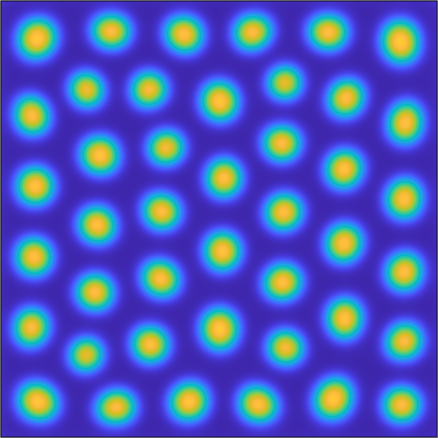}}\hspace{0.2cm}\subfloat[Mixed]{\includegraphics[width=0.23\textwidth]{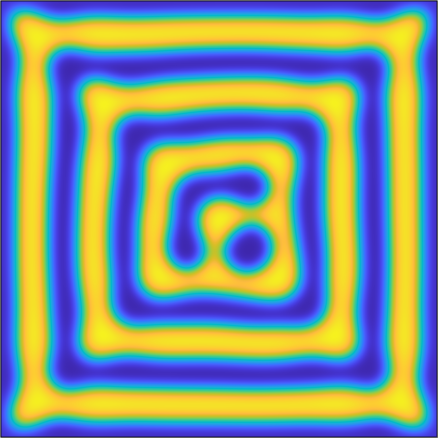}}
    
    \caption{Values of $u_1$ from simulations of \eqref{threespecs_LV} with mixed boundary conditions $u_2=1/a_{22}$, $\bm{n}\cdot \nabla u_1=\bm{n}\cdot \nabla u_3=0$ in (a)-(b), and $u_3=1/a_{33}$, $\bm{n}\cdot \nabla u_1=\bm{n}\cdot \nabla u_2=0$ in (c)-(d). We used $a_{11}=1.1$, $a_{12}=a_{13}=a_{21}=0.8$, $a_{23}=0.8$, $a_{31}=1.7$, $a_{32}=-1$, $a_{33}=1.1$, $D_1=0.1$, $D_2=100$, and $D_3=1$, with $a_{22}=1$ in (a),(c) and $a_{22}=1.3$ in (b), (d). $\Omega$ was taken to be a square domain of side length $L=300$. $u_2$ is in phase with $u_1$, and $u_3$ is out of phase with these. The colours correspond to a maximal value of $0.7$ (light yellow) and a minimal value of $0$ (dark blue). Simulations are shown at time $t=10^5$, and were initialized using normal perturbations of the homogeneous steady state as described in the beginning of the section. The colour scale ranges between $0$ (blue) and $0.7$ (yellow) in all panels. }
    \label{threespecs_fig2}
\end{figure}

We have chosen the species with a fixed Dirichlet boundary condition here to be $u_1$, which has the smallest diffusion coefficient. We now explore applying the Dirichlet condition to the other two species. As the system is multistable in the absence of diffusion, we can exploit this to choose a value of $R=1/a_{ii}$, where $i=2$ or $3$ indexes the species with the Dirichlet condition, which forces $u_1\approx 0$ in the neighborhood of the boundary (as both $(0,1/a_{22},0)$ and $(0,0,1/a_{33})$ are steady states of the kinetic part of \eqref{threespecs_LV}). We show these simulations in Figure \ref{threespecs_fig2}. In the case of $u_3$ fixed at the boundary, with the intermediate diffusion value, we see qualitatively similar solutions between Figures \ref{threespecs_fig2}(c) and \ref{threespecs_fig}(b), as well as \ref{threespecs_fig2}(d) and \ref{threespecs_fig}(d), with only small defects or pattern selection differences between them. In the case of $u_2$ being fixed at the boundary however, we see much larger regions of homogeneity extending from the boundary in Figure \ref{threespecs_fig2}(a)-(b), as the fixed value of $u_2$ can diffuse much more readily throughout the domain due to the size of its diffusive flux. This is as anticipated in the discussion at the end of Section \ref{intro_model}. We remark that taking $R=1$ has no influence on the qualitative solutions observed (not shown).

While this is only a single example of a multispecies model, we anticipate that the generalization to $n$ species will have qualitatively similar effects. Without loss of generality, the arguments given in Section \ref{asymptotic_sec} can be extended to the $n$ species case as long as all constraints are satisfied. In particular, we anticipate that placing the Dirichlet condition on any single species will confine the patterns to the interior via the same mechanism suggested in Section \eqref{pattern_selection_sec}. We do note however that interactions with multistability can lead to more complex behaviour, and may require a choice of $R$ different from what is stated in Section \ref{intro_model}. For instance, if we took $R=0$ in the examples shown in Figure \ref{threespecs_fig2}, we would see a similar localization of the interior pattern, but the region around it would have $u_1 \approx 0.7$ (i.e.~a maximal value) rather than $u_1 \approx 0$. Lastly, we remark that while this system exhibits multistability, Turing instabilities only make sense around the coexistence equilibrium of all three species (as they cannot exist around two-species coexistence equilibria, and the extinction steady state would lead to negative values of at least one species). We leave exploration of more general multistable systems to future work.

\section{Discussion}\label{discussion_section}

Isolation of patterns in reaction-diffusion systems (RDS) on the interior of a domain can be motivated both phenomenologically, by observing patterning fields which are clearly subsets of a given region, or mechanistically in terms of boundaries in gene expression. Such gene expression boundaries themselves can either emerge due to previous cell fate specification or local signalling dynamics. Here we have developed a novel method of inducing such isolated patterns with a simple change in the boundary conditions governing the RDS. We have justified these boundary conditions by considering heterogeneous RDS modelling a change in reaction dynamics explicitly, and shown that such systems can reduce asymptotically to RDS satisfying these mixed boundary conditions at steady state, given weak geometric and kinetic constraints. This is consistent with the theoretical literature showing pattern localization and modulation in heterogeneous environments \cite{varea1997confined, page2003pattern, page2005complex, krause_WKB, kozak2019pattern}, as well as the regionalization of patterning seen experimentally \cite{johansson2014regionalisation}. These conditions give a simple way of modelling these localization phenomena, without having to resort to modelling the heterogeneity that may have given rise to isolated patterning regions. We have further shown that isolated patterning can be obtained quite generically using these mixed boundary conditions across a range of reaction-diffusion systems, geometries, spatial dimensions, and even in systems of more than two interacting species. 

As described in the introduction, there are still major gaps in our understanding of RDS, particularly as models in developmental biology. While the original formulation of Turing's theory is a powerful and simple way of obtaining periodically patterned states, it suffers from a number of robustness problems, in part due to this simplicity \cite{maini2012turing, woolley2017turing, scholes2019comprehensive}. In addition to demonstrating isolated patterning, our results indicate some level of robustness in pattern selection, partly explained in Section \ref{pattern_selection_sec}. Specifically, the numerically observed steady states in the case of homogeneous Neumann conditions can generally depend on initial data, as multistability of steady states is common, especially in two or more spatial dimensions \cite{borckmans1995turing}. In contrast, there were far fewer such cases of multistability of patterns observed from simulations with the mixed boundary conditions \eqref{mixed_BCs}. Of course more work is needed to precisely explain this phenomenon, but it is consistent with the discussion of how bifurcation branches of solutions change under different boundary conditions in \cite{dillon1994pattern}. Additionally, while we did not explore this in great detail here, the Turing space for these mixed boundary conditions can be enlarged, as discussed in \cite{maini1997boundary}, plausibly helping to overcome the constraints in finding parameters which admit patterned solutions.

While we focused on analyzing terminal steady state patterns, we remark that the boundary conditions \eqref{mixed_BCs} often had transient impacts which may be relevant in applications. In particular, in every simulation investigated, we observed that these boundary conditions led to a substantially faster convergence towards the steady state pattern. In the case of labyrinthine patterns, these conditions often led to orders of magnitude faster convergence. Specifically, for the Schnakenberg kinetics with labyrinthine parameters in Table \ref{table_reactions}, the solution with conditions \eqref{mixed_BCs} was within plotting accuracy of the final steady state within $O(10^2)$ units of time, whereas the solution with homogeneous Neumann conditions only attained the final stripe orientations shown in Table \ref{table_sim} after $O(10^4)$ units of time. We conjecture that this is not, at least primarily, due to a change in linear stability of the final pattern, but is due to fewer steady states satisfying these boundary conditions, and hence less shadowing of unstable equilibria. These results are consistent with earlier results on inhomogeneous Robin conditions \cite{arcuri1986pattern}. As with the above discussion of counting equilibria, we leave further investigation to future work.

Finally, we mention that our results also demonstrate a novel application of domain geometry in robustly giving rise to patterns of qualitatively different types. Specifically, in Figure \ref{squiggle_fig} we see that inverted spots can be selected on the interior of these domains, with stripes closer to the boundaries, due to the symmetry of the domain geometry. As far as we are aware this is a novel local pattern selection mechanism quite distinct from those exploiting a reduction of dimension to transition from spots to stripes \cite{murray1981pre, murray2004mathematical}. In general the sensitivity of patterns with respect to geometry, and the use of domain geometry to influence patterning, have not been studied as extensively as other aspects of RDS, especially for more general boundary conditions.

Many periodic patterns in development are spatially localized within some region of an otherwise homogeneous tissue. There are at least three distinct ways that such regionalization can occur, described in Section \ref{DevSubsect}. Determining precisely what is happening at such boundaries is difficult, and often their exact geometry is unknown. If, for example, a periodic pattern of spots forms in one part of a field but not another, it does not immediately reveal where the functionally relevant boundary might be, whether it is graded or sharp, whether it is relevant for activator and inhibitor equally or selective for one species, and whether it promotes the formation of spots nearby (and so the boundary lies close to the edge of the spotted area) or repels them (and so it lies at some distance from the margin of the spotted area). The boundary conditions presented here give a simple way to model such phenomena, when the boundary is assumed to be repelling, without having precise knowledge of what the field is doing near or beyond the boundaries where periodic patterns are observed. The examples given in Section \ref{geometry_sec}, and in particular the `cut' domains shown in Figure \ref{cutfig}, provide ways of validating these isolating boundary conditions via experiments which alter domain geometry, as we have demonstrated strong effects of such geometry on the emergent patterns.

As mentioned in the introduction, Neumann boundary conditions can also exhibit isolated patterning if the initial data consist of internal spot solutions \cite{kolokolnikov2009spot, chen2011stability}. It would be interesting to quantify if such internal states are generic in some sense, by comparing and contrasting their basins of attraction from states with boundary spots. Such a quantification of `generic' behaviour would be valuable to inform questions of biological robustness of these mechanisms. There are also important open questions raised about pattern formation multistability with these mixed boundary conditions, and how one might choose the value of $R$ in more general systems, as discussed in Section \ref{many_species}.

There are numerous other extensions of the basic ideas here, but we restrict attention to five possible areas of future work. As above, there is work to be done in more rigorously justifying our conclusions regarding equilibria, pattern selection, and transient dynamics. As the examples in Table \ref{table_reactions} all have large diffusivity ratios $D$, the shadow-limit approaches of \cite{ward2002dynamics} (and many other references) could be used to provide further insight on interior localization. On the biological side, exploring patterning field boundaries with these kinds of models in mind could help us understand the key aspects of the detailed biological mechanisms at play. There are also important questions, raised in \cite{esposito2020open} and elsewhere, about the underlying thermodynamics of RDS, and the role that boundaries play. We also remark that there are examples of more complicated multi-domain and bulk-surface models, where the role of boundary conditions and geometry have major impacts, and there are important connections between those models and the spatially heterogeneous ones studied in Section \ref{asymptotic_sec} \cite{krause2020turing}.  Finally we have restricted our attention to studying stationary solutions, but RDS can exhibit a wide variety of spatiotemporal behaviours, such as spatiotemporal chaos and spiral waves observed in excitable systems \cite{sanchez2019turing}. Preliminary simulations of such dynamics suggests that our mixed boundary conditions can also lead to localized behaviour in these systems, with possible applications in electrophysiology and other areas (e.g.~little is known regarding how boundaries influence re-entrant wave dynamics \cite{clayton2011models}). While there has been great progress in extending Turing's insights into pattern formation and RDS, there is substantial work left, especially in terms of elucidating the connections with real embryonic development.

\begin{acknowledgements}
V.K. is grateful for support from the European Regional Development Fund-Project ‘Center for Advanced Applied Science’ (no. CZ.02.1.01/0.0/0.0/16 019/0000778) and the Mathematical Institute at the University of Oxford.
\end{acknowledgements}

\bibliographystyle{abbrv}
\bibliography{refs}

\begin{thebibliography}{10}

\bibitem{arcuri1986pattern}
P.~Arcuri and J.~Murray.
\newblock Pattern sensitivity to boundary and initial conditions in
  reaction-diffusion models.
\newblock {\em Journal of mathematical biology}, 24(2):141--165, 1986.

\bibitem{avitabile2018spot}
D.~Avitabile, V.~F. Bre\~{n}a Medina, and M.~J. Ward.
\newblock Spot dynamics in a reaction-diffusion model of plant root hair
  initiation.
\newblock {\em SIAM Journal on Applied Mathematics}, 78(1):291--319, 2018.

\bibitem{ball2001self}
P.~Ball.
\newblock {\em The Self-made Tapestry: Pattern Formation in Nature}.
\newblock Oxford University Press, 2001.

\bibitem{benson1993diffusion}
D.~L. Benson, J.~A. Sherratt, and P.~K. Maini.
\newblock Diffusion driven instability in an inhomogeneous domain.
\newblock {\em Bulletin of {M}athematical {B}iology}, 55(2):365--384, 1993.

\bibitem{borckmans1995turing}
P.~Borckmans, G.~Dewel, A.~De~Wit, and D.~Walgraef.
\newblock Turing bifurcations and pattern selection.
\newblock In {\em Chemical waves and patterns}, pages 323--363. Springer, 1995.

\bibitem{Bozzini-2015-WNA}
B.~Bozzini, G.~Gambino, D.~Lacitignola, S.~Lupo, M.~Sammartino, and I.~Sgura.
\newblock {Weakly nonlinear analysis of {T}uring patterns in a morphochemical
  model for metal growth}.
\newblock {\em Comput. Math. Appl.}, 70(8):1948--1969, 2015.

\bibitem{bradshaw2009handbook}
R.~A. Bradshaw and E.~A. Dennis.
\newblock {\em Handbook of {C}ell {S}ignaling}.
\newblock Academic press, 2009.

\bibitem{briscoe2015morphogen}
J.~Briscoe and S.~Small.
\newblock Morphogen rules: design principles of gradient-mediated embryo
  patterning.
\newblock {\em Development}, 142(23):3996--4009, 2015.

\bibitem{chen2011stability}
W.~Chen and M.~J. Ward.
\newblock The stability and dynamics of localized spot patterns in the
  two-dimensional gray--scott model.
\newblock {\em SIAM Journal on Applied Dynamical Systems}, 10(2):582--666,
  2011.

\bibitem{claxton1964determination}
J.~Claxton.
\newblock The determination of patterns with special reference to that of the
  central primary skin follicles in sheep.
\newblock {\em Journal of Theoretical Biology}, 7(2):302--317, 1964.

\bibitem{clayton2011models}
R.~Clayton, O.~Bernus, E.~Cherry, H.~Dierckx, F.~H. Fenton, L.~Mirabella, A.~V.
  Panfilov, F.~B. Sachse, G.~Seemann, and H.~Zhang.
\newblock Models of cardiac tissue electrophysiology: progress, challenges and
  open questions.
\newblock {\em Progress in biophysics and molecular biology}, 104(1-3):22--48,
  2011.

\bibitem{crampin2002pattern}
E.~J. Crampin, W.~W. Hackborn, and P.~K. Maini.
\newblock Pattern formation in reaction-diffusion models with nonuniform domain
  growth.
\newblock {\em Bulletin of {M}athematical {B}iology}, 64(4):747--769, 2002.

\bibitem{cross1993pattern}
M.~C. Cross and P.~C. Hohenberg.
\newblock Pattern formation outside of equilibrium.
\newblock {\em Reviews of modern physics}, 65(3):851, 1993.

\bibitem{de1991turing}
P.~De~Kepper, V.~Castets, E.~Dulos, and J.~Boissonade.
\newblock Turing-type chemical patterns in the chlorite-iodide-malonic acid
  reaction.
\newblock {\em Physica D: Nonlinear Phenomena}, 49(1-2):161--169, 1991.

\bibitem{de1997twist}
A.~De~Wit, P.~Borckmans, and G.~Dewel.
\newblock Twist grain boundaries in three-dimensional lamellar {T}uring
  structures.
\newblock {\em Proceedings of the National Academy of Sciences},
  94(24):12765--12768, 1997.

\bibitem{de1992three}
A.~De~Wit, G.~Dewel, P.~Borckmans, and D.~Walgraef.
\newblock Three-dimensional dissipative structures in reaction-diffusion
  systems.
\newblock {\em Physica D: Nonlinear Phenomena}, 61(1-4):289--296, 1992.

\bibitem{diego2018key}
X.~Diego, L.~Marcon, P.~M{\"u}ller, and J.~Sharpe.
\newblock Key features of {T}uring systems are determined purely by network
  topology.
\newblock {\em Physical Review X}, 8(2):021071, 2018.

\bibitem{dillon1994pattern}
R.~Dillon, P.~Maini, and H.~Othmer.
\newblock Pattern formation in generalized {T}uring systems.
\newblock {\em Journal of {M}athematical {B}iology}, 32(4):345--393, 1994.

\bibitem{duckett1994clonal}
C.~Duckett, C.~Grierson, P.~Linstead, K.~Schneider, E.~Lawson, C.~Dean,
  S.~Poethig, and K.~Roberts.
\newblock Clonal relationships and cell patterning in the root epidermis of
  arabidopsis.
\newblock {\em Development}, 120(9):2465--2474, 1994.

\bibitem{ei2013dynamics}
S.-I. Ei and T.~Ishimoto.
\newblock Dynamics and interactions of spikes on smoothly curved boundaries for
  reaction--diffusion systems in 2d.
\newblock {\em Japan Journal of Industrial and Applied Mathematics},
  30(1):69--90, 2013.

\bibitem{ermentrout1991}
B.~Ermentrout.
\newblock {Stripes or spots? {N}onlinear effects in bifurcation of
  reaction-diffusion equations on the square}.
\newblock {\em Proc. Math. Phys. Sci.}, 434(1891):413--417, 1991.

\bibitem{esposito2020open}
M.~Esposito.
\newblock Open questions on nonequilibrium thermodynamics of chemical reaction
  networks.
\newblock {\em Communications Chemistry}, 3(1):1--3, 2020.

\bibitem{falasco2018information}
G.~Falasco, R.~Rao, and M.~Esposito.
\newblock Information thermodynamics of {T}uring patterns.
\newblock {\em Physical Review Letters}, 121(10):108301, 2018.

\bibitem{fischer2006vectorial}
U.~Fischer, Y.~Ikeda, K.~Ljung, O.~Serralbo, M.~Singh, R.~Heidstra, K.~Palme,
  B.~Scheres, and M.~Grebe.
\newblock Vectorial information for arabidopsis planar polarity is mediated by
  combined aux1, ein2, and gnom activity.
\newblock {\em Current Biology}, 16(21):2143--2149, 2006.

\bibitem{fitzhugh1955mathematical}
R.~FitzHugh.
\newblock Mathematical models of threshold phenomena in the nerve membrane.
\newblock {\em The {B}ulletin of {M}athematical {B}iophysics}, 17(4):257--278,
  1955.

\bibitem{fitzhugh1961impulses}
R.~FitzHugh.
\newblock Impulses and physiological states in theoretical models of nerve
  membrane.
\newblock {\em Biophysical journal}, 1(6):445, 1961.

\bibitem{gierer1972theory}
A.~Gierer and H.~Meinhardt.
\newblock A theory of biological pattern formation.
\newblock {\em Kybernetik}, 12(1):30--39, 1972.

\bibitem{glover2017hierarchical}
J.~D. Glover, K.~L. Wells, F.~Matth{\"a}us, K.~J. Painter, W.~Ho, J.~Riddell,
  J.~A. Johansson, M.~J. Ford, C.~A. Jahoda, V.~Klika, et~al.
\newblock Hierarchical patterning modes orchestrate hair follicle
  morphogenesis.
\newblock {\em PLoS biology}, 15(7):e2002117, 2017.

\bibitem{green2015positional}
J.~B.~A. Green and J.~Sharpe.
\newblock Positional information and reaction-diffusion: two big ideas in
  developmental biology combine.
\newblock {\em Development}, 142(7):1203--1211, 2015.

\bibitem{ho2019feather}
W.~K. Ho, L.~Freem, D.~Zhao, K.~J. Painter, T.~E. Woolley, E.~A. Gaffney, M.~J.
  McGrew, A.~Tzika, M.~C. Milinkovitch, P.~Schneider, et~al.
\newblock Feather arrays are patterned by interacting signalling and cell
  density waves.
\newblock {\em PLoS biology}, 17(2):e3000132, 2019.

\bibitem{iron2000dynamics}
D.~Iron and M.~J. Ward.
\newblock The dynamics of boundary spikes for a nonlocal reaction-diffusion
  model.
\newblock {\em European Journal of Applied Mathematics}, 11(5):491--514, 2000.

\bibitem{iron2000metastable}
D.~Iron and M.~J. Ward.
\newblock A metastable spike solution for a nonlocal reaction-diffusion model.
\newblock {\em SIAM Journal on Applied Mathematics}, 60(3):778--802, 2000.

\bibitem{jensen1993subcritical}
O.~Jensen, V.~O. Pannbacker, G.~Dewel, and P.~Borckmans.
\newblock Subcritical transitions to {T}uring structures.
\newblock {\em Physics Letters A}, 179(2):91--96, 1993.

\bibitem{johansson2014regionalisation}
J.~A. Johansson and D.~J. Headon.
\newblock Regionalisation of the skin.
\newblock In {\em Seminars in cell \& developmental biology}, volume~25, pages
  3--10. Elsevier, 2014.

\bibitem{keener1998mathematical}
J.~P. Keener and J.~Sneyd.
\newblock {\em Mathematical {P}hysiology}, volume~1.
\newblock Springer, 1998.

\bibitem{kernevez1979hysteresis}
J.~Kernevez, G.~Joly, M.~Duban, B.~Bunow, and D.~Thomas.
\newblock Hysteresis, oscillations, and pattern formation in realistic
  immobilized enzyme systems.
\newblock {\em Journal of {M}athematical {B}iology}, 7(1):41--56, 1979.

\bibitem{kishimoto1985spatial}
K.~Kishimoto and H.~F. Weinberger.
\newblock The spatial homogeneity of stable equilibria of some
  reaction-diffusion systems on convex domains.
\newblock {\em Journal of Differential Equations}, 58(1):15--21, 1985.

\bibitem{klika2012influence}
V.~Klika, R.~E. Baker, D.~Headon, and E.~A. Gaffney.
\newblock The influence of receptor-mediated interactions on reaction-diffusion
  mechanisms of cellular self-organisation.
\newblock {\em Bulletin of {M}athematical {B}iology}, 74(4):935--957, 2012.

\bibitem{klika2018domain}
V.~Klika, M.~Koz\'{a}k, and E.~A. Gaffney.
\newblock Domain size driven instability: Self-organization in systems with
  advection.
\newblock {\em SIAM Journal on Applied Mathematics}, 78(5):2298--2322, 2018.

\bibitem{kolokolnikov2004bifurcation}
T.~Kolokolnikov and M.~J. Ward.
\newblock Bifurcation of spike equilibria in the near-shadow gierer-meinhardt
  model.
\newblock {\em Discrete \& Continuous Dynamical Systems-B}, 4(4):1033, 2004.

\bibitem{kolokolnikov2009spot}
T.~Kolokolnikov, M.~J. Ward, and J.~Wei.
\newblock Spot self-replication and dynamics for the schnakenburg model in a
  two-dimensional domain.
\newblock {\em Journal of nonlinear science}, 19(1):1--56, 2009.

\bibitem{kondo2010reaction}
S.~Kondo and T.~Miura.
\newblock Reaction-diffusion model as a framework for understanding biological
  pattern formation.
\newblock {\em science}, 329(5999):1616--1620, 2010.

\bibitem{kozak2019pattern}
M.~Koz{\'a}k, E.~A. Gaffney, and V.~Klika.
\newblock Pattern formation in reaction-diffusion systems with piecewise
  kinetic modulation: An example study of heterogeneous kinetics.
\newblock {\em Physical Review E}, 100(4):042220, 2019.

\bibitem{krause2019influence}
A.~L. Krause, M.~A. Ellis, and R.~A. Van~Gorder.
\newblock Influence of curvature, growth, and anisotropy on the evolution of
  {T}uring patterns on growing manifolds.
\newblock {\em Bulletin of {M}athematical {B}iology}, 81(3):759--799, 2019.

\bibitem{krause2020turing}
A.~L. Krause, V.~Klika, J.~Halatek, P.~K. Grant, T.~E. Woolley, N.~Dalchau, and
  E.~A. Gaffney.
\newblock Turing patterning in stratified domains.
\newblock {\em Bulletin of {M}athematical {B}iology}, 82:136, 2020.

\bibitem{krause_WKB}
A.~L. Krause, V.~Klika, T.~E. Woolley, and E.~A. Gaffney.
\newblock From one pattern into another: Analysis of {T}uring patterns in
  heterogeneous domains via {W}{K}{B}{J}.
\newblock {\em Journal of the Royal Society Interface}, 17:20190621, 2020.

\bibitem{kuramoto2003chemical}
Y.~Kuramoto.
\newblock {\em Chemical Oscillations, Waves, and Turbulence}.
\newblock Dover books on chemistry. Dover Publications, 2003.

\bibitem{kurowski2017two}
L.~Kurowski, A.~L. Krause, H.~Mizuguchi, P.~Grindrod, and R.~A. Van~Gorder.
\newblock Two-species migration and clustering in two-dimensional domains.
\newblock {\em Bulletin of {M}athematical {B}iology}, 79(10):2302--2333, 2017.

\bibitem{leda2009instabilities}
M.~Leda, V.~K. Vanag, and I.~R. Epstein.
\newblock Instabilities of a three-dimensional localized spot.
\newblock {\em Physical Review E}, 80(6):066204, 2009.

\bibitem{li2009dynamic}
X.~Li, A.~M. Udager, C.~Hu, X.~T. Qiao, N.~Richards, and D.~L. Gumucio.
\newblock Dynamic patterning at the pylorus: Formation of an epithelial
  intestine--stomach boundary in late fetal life.
\newblock {\em Developmental dynamics}, 238(12):3205--3217, 2009.

\bibitem{maini1997boundary}
P.~Maini and M.~Myerscough.
\newblock Boundary-driven instability.
\newblock {\em Applied Mathematics Letters}, 10(1):1--4, 1997.

\bibitem{maini2012turing}
P.~K. Maini, T.~E. Woolley, R.~E. Baker, E.~A. Gaffney, and S.~S. Lee.
\newblock Turing's model for biological pattern formation and the robustness
  problem.
\newblock {\em Interface focus}, 2(4):487--496, 2012.

\bibitem{meinhardt1983boundary}
H.~Meinhardt.
\newblock A boundary model for pattern formation in vertebrate limbs.
\newblock {\em Development}, 76(1):115--137, 1983.

\bibitem{miyamoto2005stability}
Y.~Miyamoto.
\newblock Stability of a boundary spike layer for the gierer--meinhardt system.
\newblock {\em European Journal of Applied Mathematics}, 16(4):467--491, 2005.

\bibitem{murray1981pre}
J.~D. Murray.
\newblock A pre-pattern formation mechanism for animal coat markings.
\newblock {\em Journal of Theoretical Biology}, 88(1):161--199, 1981.

\bibitem{murray2004mathematical}
J.~D. Murray.
\newblock {\em Mathematical {B}iology. {II}. Spatial models and biomedical
  applications}.
\newblock Interdisciplinary applied mathematics. Springer, New York, 2004.

\bibitem{murray2017self}
S.~M. Murray and V.~Sourjik.
\newblock Self-organization and positioning of bacterial protein clusters.
\newblock {\em Nature Physics}, 13(10):1006--1013, 2017.

\bibitem{nagorcka1985role}
B.~Nagorcka and J.~Mooney.
\newblock The role of a reaction-diffusion system in the initiation of primary
  hair follicles.
\newblock {\em Journal of theoretical biology}, 114(2):243--272, 1985.

\bibitem{nagumo1962active}
J.~Nagumo, S.~Arimoto, and S.~Yoshizawa.
\newblock An active pulse transmission line simulating nerve axon.
\newblock {\em Proceedings of the IRE}, 50(10):2061--2070, 1962.

\bibitem{nicolis1977self}
G.~Nicolis and I.~Prigogine.
\newblock {\em Self-Organization in Nonequilibrium Systems: From Dissipative
  Structures to Order Through Fluctuations}.
\newblock A Wiley-Interscience publication. Wiley, 1977.

\bibitem{page2003pattern}
K.~Page, P.~K. Maini, and N.~A.~M. Monk.
\newblock Pattern formation in spatially heterogeneous {T}uring
  reaction--diffusion models.
\newblock {\em Physica D: Nonlinear Phenomena}, 181(1-2):80--101, 2003.

\bibitem{page2005complex}
K.~M. Page, P.~K. Maini, and N.~A.~M. Monk.
\newblock Complex pattern formation in reaction--diffusion systems with
  spatially varying parameters.
\newblock {\em Physica D: Nonlinear Phenomena}, 202(1-2):95--115, 2005.

\bibitem{pearson1992pattern}
J.~E. Pearson and W.~J. Bruno.
\newblock Pattern formation in an n+ q component reaction--diffusion system.
\newblock {\em Chaos: An Interdisciplinary Journal of Nonlinear Science},
  2(4):513--524, 1992.

\bibitem{prigogine1971biological}
I.~Prigogine and G.~Nicolis.
\newblock Biological order, structure and instabilities.
\newblock {\em Quarterly Reviews of Biophysics}, 4(2-3):107--148, 1971.

\bibitem{raspopovic2014digit}
J.~Raspopovic, L.~Marcon, L.~Russo, and J.~Sharpe.
\newblock Digit patterning is controlled by a bmp-sox9-wnt {T}uring network
  modulated by morphogen gradients.
\newblock {\em Science}, 345(6196):566--570, 2014.

\bibitem{Ross2008}
J.~Ross.
\newblock {\em Thermodynamic and Stochastic Theory of Reaction--Diffusion
  Systems}, pages 41--58.
\newblock Springer Berlin Heidelberg, Berlin, Heidelberg, 2008.

\bibitem{sanchez2019turing}
F.~S{\'a}nchez-Garduno, A.~L. Krause, J.~A. Castillo, and P.~Padilla.
\newblock Turing--hopf patterns on growing domains: the torus and the sphere.
\newblock {\em Journal of theoretical biology}, 481:136--150, 2019.

\bibitem{satnoianu2000turing}
R.~A. Satnoianu, M.~Menzinger, and P.~K. Maini.
\newblock Turing instabilities in general systems.
\newblock {\em Journal of {M}athematical {B}iology}, 41(6):493--512, 2000.

\bibitem{schnakenberg1979simple}
J.~Schnakenberg.
\newblock Simple chemical reaction systems with limit cycle behaviour.
\newblock {\em Journal of theoretical biology}, 81(3):389--400, 1979.

\bibitem{scholes2019comprehensive}
N.~S. Scholes, D.~Schnoerr, M.~Isalan, and M.~P. Stumpf.
\newblock A comprehensive network atlas reveals that {T}uring patterns are
  common but not robust.
\newblock {\em Cell systems}, 9(3):243--257, 2019.

\bibitem{serna2017thermodynamic}
H.~Serna, A.~P. Mu{\~n}uzuri, and D.~Barrag{\'a}n.
\newblock Thermodynamic and morphological characterization of {T}uring patterns
  in non-isothermal reaction--diffusion systems.
\newblock {\em Physical Chemistry Chemical Physics}, 19(22):14401--14411, 2017.

\bibitem{setayeshgar1998turing}
S.~Setayeshgar and M.~Cross.
\newblock Turing instability in a boundary-fed system.
\newblock {\em Physical Review E}, 58(4):4485, 1998.

\bibitem{shoji2007most}
H.~Shoji and K.~Yamada.
\newblock Most stable patterns among three-dimensional {T}uring patterns.
\newblock {\em Japan journal of industrial and applied mathematics}, 24(1):67,
  2007.

\bibitem{smith2000roles}
D.~M. Smith, C.~Nielsen, C.~J. Tabin, and D.~J. Roberts.
\newblock Roles of bmp signaling and nkx2. 5 in patterning at the chick
  midgut-foregut boundary.
\newblock {\em Development}, 127(17):3671--3681, 2000.

\bibitem{subramanian2020pattern}
S.~Subramanian and S.~M. Murray.
\newblock Pattern selection in reaction diffusion systems.
\newblock {\em Physical Review E}, 103(1):012215, 2021.

\bibitem{taylor_patterns_2019}
N.~P. Taylor, H.~Kim, A.~L. Krause, and R.~A. Van~Gorder.
\newblock A non-local cross-diffusion model of population dynamics {I}:
  Emergent spatial and spatiotemporal patterns.
\newblock {\em Bulletin of {M}athematical {B}iology}, 82(112), 2020.

\bibitem{tucker2004cutting}
A.~Tucker and P.~Sharpe.
\newblock The cutting-edge of mammalian development; how the embryo makes
  teeth.
\newblock {\em Nature Reviews Genetics}, 5(7):499--508, 2004.

\bibitem{turing1952chemical}
A.~M. Turing.
\newblock The chemical basis of morphogenesis.
\newblock {\em Philosophical Transactions of the Royal Society of London.
  Series B, Biological Sciences}, 237(641):37--72, 1952.

\bibitem{tzou2011stationary}
J.~Tzou, A.~Bayliss, B.~Matkowsky, and V.~Volpert.
\newblock Stationary and slowly moving localised pulses in a singularly
  perturbed brusselator model.
\newblock {\em European Journal of Applied Mathematics}, 22(5):423, 2011.

\bibitem{tzou2018stability}
J.~Tzou and M.~Ward.
\newblock The stability and slow dynamics of spot patterns in the 2d
  brusselator model: The effect of open systems and heterogeneities.
\newblock {\em Physica D: Nonlinear Phenomena}, 373:13--37, 2018.

\bibitem{van_gorderthermal2020}
R.~A. Van~Gorder.
\newblock Influence of temperature on {T}uring pattern formation.
\newblock {\em Proc. R. Soc. A.}, 476:20200356, 2020.

\bibitem{van_gorder_growth_2019}
R.~A. Van~Gorder, V.~Klika, and A.~L. Krause.
\newblock Turing conditions for pattern forming systems on evolving manifolds.
\newblock {\em Journal of Mathematical Biology}, 82(1):1--61, 2021.

\bibitem{varea1997confined}
C.~Varea, J.~Arag{\'o}n, and R.~Barrio.
\newblock Confined {T}uring patterns in growing systems.
\newblock {\em Physical Review E}, 56(1):1250, 1997.

\bibitem{walsh2010extracellular}
D.~W. Walsh, C.~Godson, D.~P. Brazil, and F.~Martin.
\newblock Extracellular bmp-antagonist regulation in development and disease:
  tied up in knots.
\newblock {\em Trends in cell biology}, 20(5):244--256, 2010.

\bibitem{walton2016villification}
K.~D. Walton, M.~Whidden, {\AA}.~Kolterud, S.~K. Shoffner, M.~J. Czerwinski,
  J.~Kushwaha, N.~Parmar, D.~Chandhrasekhar, A.~M. Freddo, S.~Schnell, et~al.
\newblock Villification in the mouse: Bmp signals control intestinal villus
  patterning.
\newblock {\em Development}, 143(3):427--436, 2016.

\bibitem{ward2002dynamics}
M.~J. Ward, D.~McInerney, P.~Houston, D.~Gavaghan, and P.~Maini.
\newblock The dynamics and pinning of a spike for a reaction-diffusion system.
\newblock {\em SIAM Journal on Applied Mathematics}, 62(4):1297--1328, 2002.

\bibitem{woolley2011stochastic}
T.~E. Woolley, R.~E. Baker, E.~A. Gaffney, and P.~K. Maini.
\newblock Stochastic reaction and diffusion on growing domains: understanding
  the breakdown of robust pattern formation.
\newblock {\em Physical Review E}, 84(4):046216, 2011.

\bibitem{woolley2017turing}
T.~E. Woolley, R.~E. Baker, and P.~K. Maini.
\newblock Turing’s theory of morphogenesis: where we started, where we are
  and where we want to go.
\newblock In {\em The Incomputable}, pages 219--235. Springer, 2017.

\end{thebibliography}

 \end{document}